 \documentclass[aps,prl,amsmath,amssymb,twocolumn,showpacs, superscriptaddress,floatfix,10pt]{revtex4-1}

\usepackage{amsmath}
\usepackage{hyperref}
\usepackage{graphicx}
\usepackage{amsfonts}
\usepackage{amsthm}
\usepackage{cases}
\usepackage{bm}
\usepackage{ulem}  

\usepackage{color}
\definecolor{Blue}{rgb}{0.00, 0.00, 1.00}
\definecolor{Red}{rgb}{1.00, 0.00, 0.00}

\hypersetup{
    colorlinks=true,       
    linkcolor=red,          
    citecolor=blue,        
    filecolor=magenta,      
    urlcolor=cyan           
}

\newcommand{\nn}{\nonumber}
\newcommand{\be}{\begin{equation}}
\newcommand{\ee}{\end{equation}}
\newcommand{\bea}{\begin{eqnarray}}
\newcommand{\eea}{\end{eqnarray}}

\newcommand{\beq}{\begin{equation}}
\newcommand{\eeq}{\end{equation}}
\newcommand{\beqn}{\begin{eqnarray}}
\newcommand{\eeqn}{\end{eqnarray}}

\begin{document}

\title{Entanglement entropy growth in stochastic conformal field theory and the KPZ class}

\author{Denis Bernard}
\affiliation{Laboratoire de Physique de l'Ecole Normale Sup\'erieure, ENS, Universit\'e PSL, CNRS, Sorbonne Universit\'e, Universit\'e de Paris, 75005 Paris, France}
\email{denis.bernard@ens.fr}
\author{Pierre Le Doussal}
\affiliation{Laboratoire de Physique de l'Ecole Normale Sup\'erieure, ENS, Universit\'e PSL, CNRS, Sorbonne Universit\'e, Universit\'e de Paris,
75005 Paris, France}
\email{ledou@lpt.ens.fr}

\date{\today}

\begin{abstract}
We introduce a model of effective conformal quantum field theory in dimension $d=1+1$ coupled to stochastic noise, where Kardar-Parisi-Zhang (KPZ) class fluctuations can be observed. The analysis of the quantum dynamics of the  scaling operators reduces to the study of random trajectories in a random
environment, modeled by Brownian vector fields. We use recent results on random walks in random environments to calculate the time-dependent entanglement entropy of a subsystem interval, starting from a factorized state. We find that the fluctuations of the entropy in the large deviation regime are governed by the universal Tracy-Widom distribution. This enlarges the KPZ class, previously observed in random circuit models, to a family of interacting many body quantum systems. 
\end{abstract}

\pacs{64.60Fr,05.40,82.65Dp}


\maketitle

The growth of entanglement entropy in some quantum chaotic systems modeled
by random unitary circuits has been found \cite{NahumRC2017,Nahum2018} to exhibit features of the Kardar-Parisi-Zhang (KPZ) universality class. The KPZ class, which contains the continuum KPZ equation \cite{KPZ} modeling the classical stochastic growth of an interface, or equivalently the free energy of
continuum directed paths in a random potential \cite{kardareplica}, is characterized in $d=1$ by super-diffusive dynamics \cite{exponent,Johansson2000,HalpinReview},
$x \sim t^{2/3}$, and by universal Tracy-Widom distributions \cite{TW1994} related to random matrix theory
\cite{spohn2000,png,KrugReview,we-flat,corwinsmallreview,QuastelSpohnReview2015}.
It contains a number of solvable discrete classical stochastic models, of e.g.
1d particle transport, such as the asymmetric simple exclusion process (ASEP) \cite{DerridaASEP}, and directed polymers in random $d=1+1$ media \cite{Johansson2000}.
Various KPZ class behaviors was also  recently claimed in other quantum models, chaotic \cite{NahumOTOC2017} or integrable \cite{ProsenHeisenbergKPZ,deNardisAnomalous2019,AnomalousXXZ2019}, as well as in classical models, integrable \cite{ProsenIntegrableKPZ,IntegrableSpinChain2019,ClassicalXXZ2019} or not \cite{SpohnMiniReview2019,Kuramoto2019}, with no obvious connections.

To which extent KPZ-like behaviors are universal for information spreading in noisy or chaotic many-body quantum systems is still unclear. In fact, apart from the random quantum circuits (in the limit of large on site dimension), most of the evidence is numerical, and a bona-fide derivation of KPZ behavior has been elusive. However, the model we are going to present yields extra supports for the robustness of such KPZ behaviors.

Recently, the class of quantum circuits has been extended to include solvable
models \cite{BertiniProsen2019,BertiniKosProsenShort} which show spectral form factor growths typical of chaotic systems but with operator spreading concentrated along the light-cone. This points towards a possible connection between simple extended chaotic systems and, possibly noisy, conformal field theory (CFT). 

In the present paper, we show that KPZ class behaviors emerge in yet another class of 1d stochastic quantum models. Specifically it manifests itself in the large deviations of certain time-dependent correlations, among which the entanglement entropy, and hence in particular in the entanglement entropy growth.
Our results rely on using an exact representation of the quantum correlations in terms of classical diffusions in time-dependent random fields and the recently discovered connection \cite{BarraquandCorwinBeta,TTPLDBeta,CorwinGu,TTPLDDiffusion,BarraquandSticky,BarraquandPLD2019} between such diffusion problems and the KPZ equation.

The models we consider are certain stochastic perturbations of CFT, which can be viewed as CFT in random geometry.
They code for stochastic quantum dynamics as random quantum circuits do. 
They are continuous space analogs of models of spin chain submitted to stochastic baths, which were shown to provide quantum extensions of the simple symmetric exclusion process (SSEP) \cite{BernardJinQSSEP2019}.  KPZ class behaviors was indeed claimed numerically in such stochastic spin chains~\cite{KnapNoisySpin2018}.  Quantum extensions of the ASEP are known to be obtainable \cite{BernardJinKrajenbink} by promoting the noise to quantum noise \cite{QstoNoise}. These stochastic models code for fermions hopping along the chain with noisy amplitudes.  
 Our stochastic CFT are thus defined by coupling an external noise to the energy-momentum tensor components, which generate left/right chiral moves within the CFT. As a consequence, operator spreading is predominantly concentrated close to the light-cone, except for rare, but important, events.

Note that slightly different versions of stochastic CFTs was considered in \cite{BernardDoyonStochasticCFT}
to model elastic scattering in a classically fluctuating environments \cite{footnote0}: a crossover from
ballistic motion to diffusion and localization was found (see also 
\cite{LangmannMoosavi} for a static version). 

We consider a quantum conformal field theory (CFT) in dimension $d=1+1$ viewed as the low
energy effective field theory of a gapless many body system. The low energy states span
a Hilbert space. It is equipped with its two component energy-momentum tensor $T(x)$
and $\bar T(x)$, $x \in \mathbb{R}$ such that their sum $h_0(x) = v(T(x) + \bar T(x))$ is the energy density operator
and the difference $p(x) = T(x) - \bar T(x)$ is the momentum density operator. In (unperturbed and noiseless)
CFT, the dynamics is generated by a Hamiltonian $H_0 = \int dx\, h_0(x)$ and the unitary evolution on the
Hilbert space is described by the operator ${\cal U}^0_t=e^{- i t H_0}$. We now couple this system to
space time dependent noise and define the flow between time $t$ and $t+dt$ of the perturbed unitary evolution
as
${\cal U}_{t+dt} {\cal U}^\dagger_t = e^{- i dH_t}$
with Hamiltonian increment
\be \label{dHt} 
dH_t := H_0 dt +  \int dx \, (dW^+_t(x) T(x) + dW^-_t(x) \bar T(x)) ,
\ee 
where $W_t^\pm(x)$ are two space dependent stochastic processes.
We choose their increments to be of the form
\be \label{eq2} 
dW^\pm_t(x) := \xi^\pm_t(x) dt + \sqrt{D_0}\, dB^\pm_t ,
\ee
where $B^\pm_t$ are two independent standard (spatially homogeneous
\cite{footnote6}) unit Brownian 
motions, $D_0$ the bare diffusion coefficient. The two random fields $\xi^\pm_t(x)$ are 
centered gaussian space time white noise with covariance
$\mathbb{E}[\xi^\epsilon_t(x) \xi^{\epsilon'}_s(y)] = \kappa \delta_{\epsilon,\epsilon'}  \delta_a(x-y) \delta(t-s)$,
where $a$ is a short distance cutoff, and $\delta_a(x)$ is a 
mollifier of the delta function. Both $a$ and the Brownian with $D_0>0$ regularize the
trajectories, as in turbulent transport \cite{GawedzkiHorvai2004,BernardGawedzkiKupiainen,LeJanRaimond}. Here we are interested in performing averages over $B^\pm$ and 
studying sample to sample fluctuations with respect to the realisations of the
random fields $\xi^\pm_t(x)$.

\begin{figure}
\includegraphics[width=0.55\linewidth]{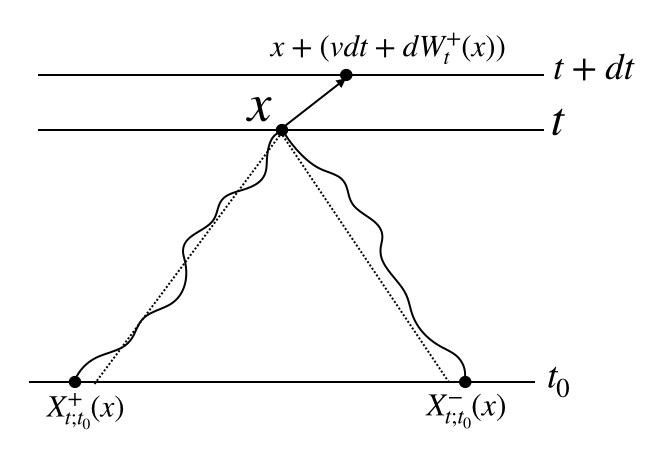}
{\caption{Top: Typical trajectories for the diffusions 
described by Eq. \eqref{langevin} in the random fields $\pm (v + \dot W^{\pm}(x))$
starting at $X^\pm_{t;t_0}(x)$ at time $t_0$ and ending at $x$ at time $t$. (The dotted lines are eye guides to visualise the light-cones, in the absence of random fields). The top of the figure
illustrates how Eq. (5) is obtained by extending the trajectory (which is at $x$ at time $t$)
from time $t$ to $t+dt$.}
\label{fig:X+X-}}  
\end{figure}

We now show how to solve for the equations of motion to make the link with trajectories in random environments.
Since $T(x)$ and $\bar T(x)$ are generators of diffeomorphism on $\mathbb{R}$,
the coupling \eqref{dHt} ensures that the dynamics of the chiral operators
is linked to the trajectories associated to the random vector fields $W_t^\pm(x)$.
In CFT it is sufficient to look at the primary operators with well defined scaling
dimensions. These operators appear in two classes depending on their commutation
relation with $T$ and $\bar T$. In pure CFT, they are convected along the
light cone with velocities $\pm v$ (right/left movers). 
As shown in \cite{SM}, the evolution of any chiral operator $\varphi(x)$ of
dimension $\Delta$ is solved as
\bea \label{evolution} 
&& \varphi(x,t) := {\cal U}_t^\dagger \varphi(x) {\cal U}_t = [X_t^{+\prime}(x)]^\Delta \varphi(X^+_t(x)) ,\\
&& \bar \varphi(x,t) := {\cal U}_t^\dagger \bar \varphi(x) {\cal U}_t = [X_t^{-\prime}(x)]^\Delta  \bar \varphi(X^-_t(x)) ,
\eea 
where the prime denotes derivative with respect to $x$. Here 
$X^\pm_t(x)$ are the processes specified by
\be \label{5} 
X_{t+dt}^\pm(x \pm v dt \pm dW^\pm_t(x)) = X_t^\pm(x) .
\ee
Eqs.\eqref{5} have a natural interpretation in terms of random trajectories.
Consider the following problem of diffusion of a particle in a random field,
whose position $x_u^+$ (resp. $x_u^-$) at time $u \in [t_0,t]$ obeys
the Langevin equation 
\be \label{langevin} 
\frac{d x_u^\pm}{du}  = \pm \big( v + \xi^\pm_u(x_u^\pm) + \sqrt{D_0}\, \frac{d B_u^\pm}{du} \big) .
\ee
From a geometrical perspective, these two equations are those of the null geodesics in the random metric associated to these stochastic fields (see \eqref{eq:Umetric} in \cite{SM}). 
It is clear that $X^\pm_t(x)$ is then the position at the initial time $t_0=0$ of the particle 
which will be at position $x$ at time $t$, and is associated to a trajectory $x^\pm_u$ such that
$x_{u=t_0}^\pm=X_t^\pm(x)$ and $x_{u=t}^\pm=x$. See Figure \ref{fig:X+X-}.

The above result allows to express time-dependent correlation functions of primary fields 
(chiral or anti-chiral)
at points $x_i$ and time $t$, in terms of the initial correlations but at positions transported
by the backward flow, i.e. at positions $X^\pm_t(x_i)$. Statistical properties of these quantum
correlations reduce to those of the random trajectories $x^\pm_{i;u}$.

We now use this property to calculate the Renyi entanglement entropy for an interval $[x_0,x]$ on the real axis.
It is defined as $S_n = \frac{1}{1-n} \log {\rm Tr} \rho_{x_0,x}(t)^n$ where
$\rho_{x_0,x}(t)$ is the reduced density matrix at time $t$, obtained by
tracing out the degrees of freedom in the domain complementary to the interval $[x_0,x]$.
Within field theory, this Renyi entropy can be represented by a path integral
on a $n$-sheet branched covering of the spacetime plane \cite{Cardy1}. It leads to 
express $S_n$ in terms of quantum correlations as
\be
{\rm Tr} \rho_{x_0,x}(t)^n= a_0^{4 \Delta_n} \langle \Psi_0 | \tilde \Phi_n(x_0,t) \Phi_n(x,t) |\Psi_0 \rangle ,
\ee 
where $\Phi_n(x,t)$ and $\tilde \Phi_n(x,t)$ are the time evolved
of the so-called conjugated twist operators, $\Phi_n(x)$ and $\tilde \Phi_n(x)$,
which implement the permutations of the sheets of the covering space at the branching points
\cite{footnote01}. Here
$|\Psi_0\rangle$ is the initial state at $t_0=0$ of the full domain (here the real axis)
and $a_0$ is an UV cutoff length. 
Decomposing the twist operators in chiral and anti-chiral components
as $\Phi_n(x)=\varphi_n(x) \bar \varphi_n(x)$, and using the formula \eqref{evolution} for the 
primary fields $\varphi_n$ and $\bar \varphi_n$, one obtains 
\be \label{Sn} 
e^{(1-n) S_n} = J_t(x_0,x)^{\Delta_n} G_t(x_0,x) ,
\ee
where $J_t(x_0,x)= X^{-\prime}_t(x_0) X^{+\prime}_t(x_0) X^{-\prime}_t(x) X^{+\prime}_t(x)$ is the Jacobian
associated to the stochastic flows \eqref{langevin}, and 
$G_t(x_0,x)$ is equal to the following quantum correlation 
\be
\langle \Psi_0 | \varphi_n(X^-_t(x_0)) \bar \varphi_n(X^+_t(x_0)) \varphi_n(X^-_t(x)) \bar \varphi_n(X^+_t(x)) |\Psi_0 \rangle . 
\ee 
The scaling dimension of the twist operators is $\Delta_n= \frac{c}{12} (n- \frac{1}{n})$, with $c$ the CFT central charge. 

We must now specify the initial state $|\Psi_0\rangle$. As in random circuit models and in
quantum quench problems, we choose a gapped initial state with finite and
small coherence length $v \tau_0$. This mimics a fully factorized state on the lattice. 
Within CFT we use the Calabrese-Cardy representation of such a state as
\cite{CalabreseCardy1,CalabreseCardy4}
\be
| \Psi_0 \rangle \propto e^{- \tau_0 H_0} |B \rangle \rangle ,
\ee 
where $|B \rangle \rangle$ is a (unnormalized) conformally invariant boundary state.
This representation comes with rules for calculation of expectation values which
involve analytic continuation of CFT correlation functions in a strip of width $2 v \tau_0$
conformally mapped to the upper half plane. Using these rules (see \cite{SM})
one obtains
\be \label{Gt} 
G_t(x_0,x)=(\frac{\pi a_0}{2 v \tau_0})^{4 \Delta_n} [ \frac{(z_0^+ z_0^- z^+ z^-)}{(z_0^+ - z^+)^2(z_0^--z^-)^2}]^{\Delta_n} K(\eta) ,
\ee
where $z^\pm = \pm i e^{\frac{\pi}{2 v \tau_0} X^\pm_t(x)}$ and $z_0^\pm = \pm i e^{\frac{\pi}{2 v \tau_0} X^\pm_t(x_0)}$.
The variable $\eta$ is the cross-ratio $\eta= \frac{(z_0^+-z_0^-)(z^+-z^-)}{(z_0^+-z^-)(z^+-z_0^-)}$
and $K(\eta)$ is called a conformal block. The above expression is in general difficult to evaluate
but it simplifies in the limit of a large interval $x_0 \to - \infty$ with fixed endpoint $x$, because one can use
the operator product expansion (OPE) in that limit. Indeed, we expect that
$X_t(x_0) \to -\infty$ in that limit a.s. for any fixed $t$. Thus
$z_0^\pm$ tend to zero along the imaginary axis,
hence $\eta \ll 1$. One knows, from boundary OPE in CFT, the asymptotics of $K(\eta) \simeq A_\phi^2
 \eta^{- (2 \Delta_n-\Delta_b)}$ as $\eta$ goes to zero, where $\Delta_b$ is the scaling dimension of the
 boundary operator produced by the OPE and $A_\phi$ is a universal amplitude.
 From Ref. \cite{CalabreseCardy1,CalabreseCardy4}
 one has $\Delta_b=0$. 
 In this limit \eqref{Gt} becomes, to leading order,
 \bea \label{Gt2} 
 && G_t(x_0,x) \simeq  \\
 && 
 \frac{A_\phi^2 (\frac{\pi a_0}{4 v \tau_0})^{4 \Delta_n} }{[ \cosh( \frac{\pi (X_t^+(x_0)-X_t^-(x_0))}{4 v \tau_0} )
 \cosh( \frac{\pi (X_t^+(x)-X_t^-(x))}{4 v \tau_0})  ]^{2 \Delta_n}} . \nonumber
 \eea 
 In the limit where the coherence length $v \tau_0$ of the initial state $|\Psi_0 \rangle$ is small, 
 the non-vanishing of this correlation function conditions the two trajectories to start at
 nearby positions $X_t^+(x) \approx X_t^-(x)$. Estimating the probability of this event will 
 thus be of importance below.
 
We are interested in various averages of the entropy $S_n$ over the Brownian, for a fixed random field configuration $\xi_t^\pm(x)$
(i.e. fixed sample). Convenient averages have the form $- \frac{1}{q} \log 
\langle e^{-q S_n } \rangle_B$ as a function of the parameter $q$ varying from annealed average for $q=1$ 
to the quenched average $\langle S_n \rangle_B$ for $q \to 0$. 
Let us take the power $q/(n-1)$ of \eqref{Sn}, and average over the Brownian.
Neglecting the terms containing $x_0$ in the limit $x_0 \to - \infty$ 
\cite{footnote3} and 
setting for now the Jacobian factor to unity, we obtain
\be \label{eq} 
\langle e^{-q S_n } \rangle_B \simeq e^{- q s_n}
\int dy^+ dy^- 
\frac{P^+_{t,t_0}(x,y^+) P^-_{t,t_0}(x,y^-) }{\cosh(\frac{\pi (y^+-y^-)}{4 v \tau_0} )^{2 q \delta_n}} ,
\ee 
where $\delta_n = \frac{\Delta_n}{n-1}= \frac{c}{12} \frac{n+1}{n}$
and $s_n$ is the (non-universal) initial value of the entropy. 
We have introduced the following probability distribution function (PDF)
\be
P^\pm_{t,t_0}(x,y) = \langle \delta(X^\pm_{t,t_0}(x) - y) \rangle_B .
\ee 
Here, to decipher the variation in the initial time $t_0$, we denote more explicitly $X^\pm_{t,t_0}(x):=X_t(x)$ the position at time $u=t_0$ of the particle 
diffusing as in \eqref{langevin} which will be at position $x$ at time $u=t$.
Thus $P^\pm_{t,t_0}(x,y)$ can be interpreted as the probability that the time reversed path 
starting from $x$ at $t$ ends at $y$ at $t_0$. It
satisfies the Fokker-Planck equation as $t_0$ is decreased,
\be \label{FP}
- \partial_{t_0} P^\pm_{t,t_0}(x,y)  = [\frac{D}{2} \partial_y^2 \pm \partial_y (v+\xi^\pm_{t_0}(y) ) ]
P^\pm_{t,t_0}(x,y) ,
\ee 
in the time reversed field $\mp (v + \xi^\pm)$, with the condition $P^\pm_{t,t}(x,y)=\delta(x-y)$,
and where $D$ is the coarse-grained diffusion coefficient
(see \cite{SM} for details). Note that, as shown in \cite{SM}, Eqs. \eqref{eq} and \eqref{FP} 
can be extended to include the contributions of the Jacobian factor in Eq. \eqref{Sn}, however
the latter are irrelevant at large scale and only renormalize a few amplitudes as
detailed in \cite{SM}.

The problem of diffusion in a time dependent random field was recently 
found to be related to the KPZ class. This was shown for discrete
random walks models in a time dependent random environment, through exact solutions \cite{BarraquandCorwinBeta,TTPLDBeta} and in their weak disorder/continuum limit
\cite{CorwinGu}. The continuum model was studied in 
\cite{TTPLDDiffusion} using physics arguments, and rigorously recently \cite{BarraquandSticky} (although the
space-time white noise limit for $\xi^\pm$ remains mathematically challenging). 
The main idea is as follows. For $\xi^\pm$ white noise in time, the disorder average  
\be
\mathbb{E}[ P^\pm_{t,t_0}(x,y) ] = \frac{1}{\sqrt{2 \pi D t}} \,e^{- \frac{(y-x \pm v (t-t_0))^2}{2 D (t-t_0)}} 
\ee 
is identical to the pure biased diffusion, i.e. $\xi=0$, and gives the global shape of the PDF,
which is centered around $y=x \mp v (t-t_0)$. The KPZ physics arises {\it away} from this
most probable direction, i.e. for $y - x \sim (\mp v + \theta_\pm) t$ at large $t$ (for $t_0$ fixed)
with $\theta_\pm \neq 0$.
For $y-x = o(t)$, i.e. $\theta_\pm= \pm v$, the average profile at large time varies in space as $e^{\mp \frac{(y-x) v}{2 D}}$.
Defining the fluctuation field around the average profile as 
$Z^\pm_{t,t_0}(x,y):=P^\pm_{t,t_0}(x,y) e^{\pm \frac{(y-x) v}{D} + \frac{v^2}{2 D} (t-t_0)}$,
one finds that $Z^\pm$ satisfies the stochastic heat equation (SHE) as $t_0$ is decreased, $- \partial_{t_0} Z^\pm  = [\frac{D}{2} \partial_y^2 - \frac{v}{D}\xi^\pm_{t_0}(y) ] Z^\pm + \dots$,
where the additional term $\dots = \partial_y (\xi Z)$ can be argued to be irrelevant at large time 
\cite{TTPLDDiffusion} in the region $y-x = o(t)$ (since it contains higher gradients).
This property is supported by recent results \cite{BarraquandSticky,BarraquandPLD2019}. Here $Z$ can be seen as the partition sum of a directed
polymer in the time dependent random potential $\frac{v}{D}\xi^\pm_{t}(y)$ and 
$h^\pm = \log Z^\pm$ is the corresponding KPZ height field \cite{SM}. 
From known results on the KPZ equation, one expects, at large time $t$, universal 
height fluctuations w.r.t. $\xi$, scaling as $\delta h \sim t^{1/3}$, with space-time scaling $x \sim t^{2/3}$.
If the region $y-x = o(t)$ dominate the integral we can thus write
\be \label{eq2} 
\! \! \langle e^{-q S_n } \rangle_B \sim  e^{-\frac{v^2}{D}(t-t_0)}
\int \! \! dy^+ dy^- 
\frac{Z^+_{t,t_0}(x,y^+) Z^-_{t,t_0}(x,y^-) }{\cosh(\frac{\pi (y^+-y^-)}{4 v \tau_0} )^{2 q \delta_n}} .
\ee 
\begin{figure}
\includegraphics[width=0.8\linewidth]{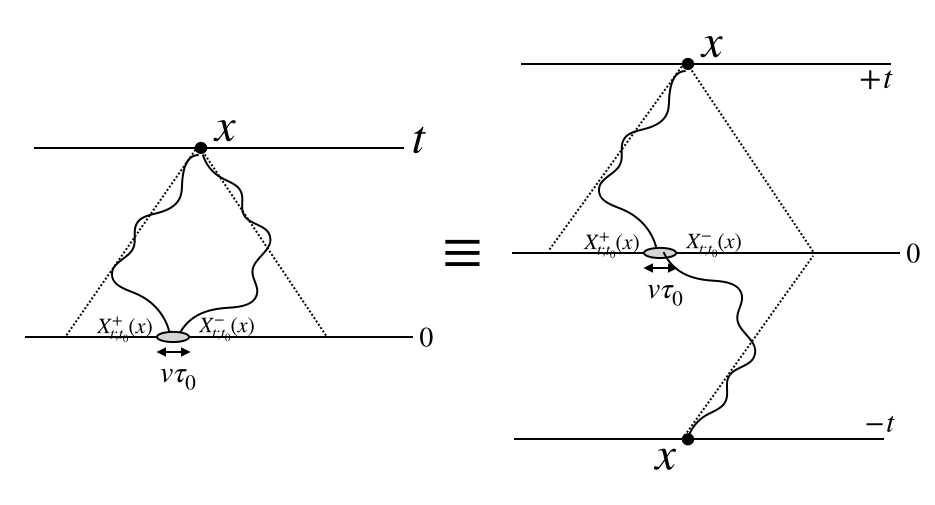}
{\caption{Left: Atypical trajectories contributing to \eqref{eq}, i.e. constrained
to be at almost identical positions at time $t_0=0$ (conditioned to be both at $x$ at time $t$).
Right: equivalent unfolding of the left picture on a time interval $2t$, with one
of the two trajectories time-reversed. As shown in the text, this geometrical configuration
applies for $q>q_c$, while for $q<q_c$ the two relevant trajectories do not meet at time $t_0=0$.}
\label{fig:X+X-Merges}}  
\end{figure}
At large time the typical KPZ spatial scale is
$t^{2/3} \gg v \tau_0$, hence
for fixed $q>0$ one can 
approximate in both Eq. \eqref{eq2} and \eqref{eq} the factor 
$1/\cosh(\frac{\pi (y^+-y^-)}{4 v \tau_0} )^{2 q \delta_n} \sim v \tau_0 \delta(y^+-y^-)$,
which constrains the two trajectories $x^\pm_s$ in \eqref{langevin}
to be in an atypical configuration with near identical
starting and ending points. Since 
$\xi^+$ and $\xi^-$ are uncorrelated fields, this allows 
to unfold the configuration by time reversing one of the two
trajectories, see Figure \ref{fig:X+X-Merges}. The problem becomes equivalent
in law, to computing the probability density of return $P^+_{2t,0}(x,x)$
to the starting point $x$,
after time lapse $2 t$, in a random vector field with constant bias $v$.
Hence,
\be \label{eq3} 
\! \! \! \langle e^{-q (S_n-s_n) } \rangle_B \simeq v \tau_0 \,  
P^+_{2t,0}(x,x) = v \tau_0 \, Z^+_{2t,0}(x,x) e^{-\frac{v^2}{D} t} ,
\ee 
similar to the point to point directed polymer partition function. From known
results and universality in the KPZ class \cite{QuastelKPZFP},
we thus obtain that at large time
\be \label{res1} 
\log \langle e^{-q (S_n-s_n) }  \rangle_B \simeq - c_1 t + c_2 t^{1/3} {\cal A}(\hat x, \hat x)\,, \quad 
\hat x = \frac{x}{2 c_3 t^{2/3}},
\ee 
where $c_1,c_2,c_3$ are constants independent of $q$ and $n$,
and ${\cal A}(\hat x,\hat y)$ denotes the so-called
Airy sheet process \cite{QuastelAirySheet,ViragAiSheet,BorodinShift}. We recall that, for fixed $\hat y$, it is equal to
${\cal A}(\hat x,\hat y) \equiv {\cal A}_2(\hat x - \hat y) - (\hat x-\hat y)^2$ where
${\cal A}_2(\hat x)$ is the Airy$_2$ process \cite{PrahoferSpohn} (describing, upon rescaling, the rightmost particle of the
Dyson Brownian motion \cite{SpohnProlhac}). Its one point PDF for $\chi = {\cal A}(\hat x, \hat x) \equiv_{\rm in law} {\cal A}_2(0)$
is given by the GUE Tracy Widom distribution \cite{TW}. Hence, we conclude
that, at fixed $q$, $\log \langle e^{-q S_n }  \rangle_B$ is distributed 
according to GUE-TW. The $x$ dependence however
of \eqref{res1}, i.e. with $S_n \equiv S_n(x,t)$, is non-trivial
and related to the Airy sheet.

Let us discuss now the PDF of $S_n$. Taking the logarithm in 
\eqref{Sn} and \eqref{Gt2} we have
\bea \label{typ} 
&& S_n = s_n + 2 \delta_n \log \cosh(\frac{\pi (X_t^+(x)-X_t^-(x))}{4 v \tau_0}) \\
&& \simeq s_n + 2 \delta_n \frac{\pi |X_t^+(x)-X_t^-(x)|}{4 v \tau_0} .
\eea 
For any given $x$,  $X_t^+(x)-X_t^-(x)$ has the same statistics as a diffusion in a biased random field
for time duration $2t$. Indeed, since $\xi^+$ and $\xi^-$ are independent
vector fields, we can reflect one of them around the space slice at
time $t$ to define a gaussian white noise vector field on a doubled time interval.
As a consequence, $|X_t^+(x)-X_t^-(x)| \equiv |\tilde x_{2t} -\tilde x_0|$
where $\tilde x_s$ are the corresponding trajectories (see \cite{SM} for detail). 

Let us now focus on the typical behavior of $S_n$.
We can use
arguments and exact results from the diffusion problem. The typical
value of $x_{2t}$ is $2 v t$ with variance $2 t D$, hence 
the typical value of $S_n$ is $S^{\rm typ}_n= \pi \delta_n t/ \tau_0$
with a variance $\mathbb{E}[\langle S_n^2 \rangle_B]^c= D_s t$ 
with $D_s=(\frac{\delta_n \pi}{2 v \tau_0})^2 2 D$
(note that the effect of the Jacobian is to renormalize $D_s$ \cite{SM}).
As a function of the end point position $x$, when 
varied over regions $x=O(t^{1/2})$, $S_n(x,t)$ 
exhibits subleading sample to sample fluctuations
$O(t^{1/4})$ described by the Edwards-Wilkinson equation (i.e. the KPZ equation
without the non-linear term) \cite{SM}.

For atypical fluctuations, the large deviations of $S_n$ can be obtained
from the large deviations of $\tilde x_{2t} -\tilde x_0$. These have been studied 
in the context of diffusion in time dependent random environments
\cite{BarraquandCorwinBeta,BarraquandSticky,TTPLDDiffusion,BarraquandPLD2019}
and one obtains at large time \cite{footnote4}
(see \cite{SM}) for $\theta \geq - v$
\bea \label{ld} 
&&  \hskip -1.6cm  \log {\rm Prob}(S_n >  \frac{\pi \delta_n}{v \tau_0}
(\theta+ v) t)   \nn  \\
&&  \simeq -  J(\frac{\theta}{\theta^*}) \, \frac{2 t}{t^*} \, + (\frac{2 t}{t^*} )^{1/3} \, G(\frac{\theta}{\theta^*}) \, \chi_2 .
\eea
Here $t^* = \kappa^2/D^3$ and $\theta^* = D^2/\kappa$ are characteristic
scales of the diffusion in a Brownian field. Note that at small $\theta$ one has
$J(\theta) = \frac{\theta^2}{2} + O(\theta^4)$ and $G(\theta)\simeq \frac{1}{2^{1/3}}\theta^{4/3}(1+ O(\theta^2))$ 
from the universal KPZ regime \cite{SM}.

One now asks how the result \eqref{ld} for the probability matches the result \eqref{res1}
for the exponential moments. This will allow to specify the domain of validity
of \eqref{res1} as a function of $q$ for a small but finite $\tau_0$. It shows that there
is a phase transition at a critical value $q=q_c$ which corresponds to a change
in the geometry of the contributing atypical trajectories.
At large time we can
evaluate the integral on $\theta$ representing the exponential moments using
a saddle point method, which gives the estimate
\bea \label{col} 
&& \log \langle e^{-q (S_n-s_n) }  \rangle_B \\
&& \simeq
- \min_{\theta} \left[ q \frac{\pi \delta_n}{2 v \tau_0} |\theta+v| 2 t + J(\frac{\theta}{\theta^*}) \, \frac{2 t}{t^*} 
- (\frac{2 t}{t^*} )^{1/3} \, G(\frac{\theta}{\theta^*}) \chi_2 \right] \nn
\eea 
There is a transition as a function of $q$ in this minimisation problem. There exists a $q_c>0$
such that for $q>q_c$ the minimum is frozen at $\theta=-v$ independant of $q$, 
in which case formula \eqref{res1} holds, with 
$c_1=(2/t^*)J(-v/\theta^*)$ and $c_2=(2/t^*)^{1/3}G(-v/\theta^*)$. For $q<q_c$, the maximum is at $\theta=\theta_c(q)>-v$. 
Let us estimate $q_c$ in the case $v < \theta^*=D^2/\kappa$, where we can use the quadratic approximation
for $J(\theta)$. One finds that $\theta_c(q)=- \frac{q}{q_c} v$  where $q_c \simeq \frac{2 v^2 \tau_0}{D \pi \delta_n}$ \cite{footnote5}, and, instead of \eqref{res1}, we obtain the leading
behavior at large time $\log \langle e^{-q (S_n-s_n) }  \rangle_B = - \frac{v^2 t}{D} \frac{q}{q_c}(2-\frac{q}{q_c}) + o(t)$, where we observe that the small $q$ expansion yields 
the first two cumulants of $S_n$ compatible with the analysis above based on
typical events \cite{SM}. One can show \cite{SM} that the $o(t)$ term exhibits sample to sample fluctuations, described
by the sum of two TW-GOE distributions. 

This saddle point estimate has a nice geometrical interpretation
in terms of trajectory configurations, For $q>q_c$ the optimal paths join at time $t_0=0$ in the
center $y=x$ up to $o(t)$ fluctuations (of KPZ type of order $O(t^{2/3})$).
For $q<q_c$ the $\pm$ optimal paths are separated by an angle $2 (v+\theta_c) \simeq 2(1-\frac{q}{q_c}) v$ in the
$(x,t)$ plane.

The above considerations can be extended to predict the fluctuations of other observables \cite{footnote6}. For instance one can calculate the time 
evolved one point function $\langle \Psi_0| \Phi(x,t) | \Psi_0 \rangle$. In the absence of randomness this 
expectation value is known to decrease exponentially with time \cite{CalabreseCardy1,CalabreseCardy2,CalabreseCardy3}.
In presence of randomness a calculation similar to the above gives 
\be \label{single}
\langle \Psi_0| \Phi(x,t) | \Psi_0 \rangle = 
 \frac{C_\Phi (\frac{\pi a_0}{4 v \tau_0})^{2 \Delta} [X^{-\prime}_t(x) X^{+\prime}_t(x)]^{\Delta}  }{
 [\cosh( \frac{\pi (X_t^+(x)-X_t^-(x))}{4 v \tau_0})  ]^{2 \Delta}} ,
\ee
where $2 \Delta$ is the scaling dimension of the primary operator $\Phi$, and $C_\Phi$ is a non-universal constant.
One similarly finds that the typical fluctuations behave as
\be
\log \langle \Psi_0| \Phi(x,t) | \Psi_0 \rangle \simeq -  \pi \Delta \frac{t}{\tau_0} + \sqrt{D_\Delta} B(t) ,
\ee 
with $B(t)$ a unit Brownian, 
i.e. the typical decay is exponential, with log-normal
fluctuations. The large deviation function of the 
atypical fluctuations are controlled by the KPZ class,
similarly as for the entropy. Others examples are 
the correlations of the components $T$ and $\bar T$
of the energy momentum tensor, which can be calculated
because each of the components is transported covariantly by the
random flows (see \eqref{TT} in \cite{SM}). For example, the connected
two point function of $T$ 
\bea \label{TT0}
&& \langle \Psi_0|T(x_1,t_1) T(x_2,t_2) |\Psi_0 \rangle^c  
\\
&& = (X^{+\prime}_{t_1}(x_1))^2 (X^{+\prime}_{t_2}(x_2))^2 \frac{c/2}{(\cosh( \frac{\pi}{4 v \tau_0} (X^+_{t_2}(x_2) 
- X^+_{t_1}(x_1)))^4} \nonumber 
\eea 
As detailed in \cite{SM} the analysis of the trajectories allows to exhibit KPZ type
large deviations. 

In conclusion, we have introduced a stochastic version of $d=1+1$ CFT modeling random unitary dynamics in interacting many body systems. We have analysed the statistical behaviors (the typical and atypical behaviors) of various quantum correlations, including the entanglement entropy of a subsystem. Geometrically, the rare events we analysed correspond to null geodesics converging to nearby points, as caustics do. Within stochastic CFT, these behaviors are universal in the sense that they only rely on the conformal symmetry acting on the physical Hilbert space of the model. We have been able to decipher them by mapping their analysis to that of random trajectories in random fields. For systems initially prepared in short range correlated states, we found that the large deviations of the fluctuations of the entanglement entropy is controlled by the KPZ class universality. 
The mechanism for the emergence of KPZ behavior appears to be different from the one unveiled in
the studies of random quantum circuits. To understand the extent by which such KPZ-like behaviors are universal for information or operator spreading in noisy or chaotic many body quantum systems remains an important question.


{\it Acknowledgments:} We especially thank G. Barraquand for very helpful discussions.
We are also grateful to B. Doyon and A. Nahum, for enlightening discussions.
PLD acknowledges support from ANR under the grant 
ANR-17-CE30-0027-01 RaMaTraF. 

{}

\newpage
~\vfill \eject

\onecolumngrid 

\begin{widetext}

\begin{large}
\begin{center}
{\bf Supplementary Material for\\ ~\\
 {\it Entanglement entropy growth in stochastic conformal field theory and the KPZ class} }
\end{center}
\end{large}

\bigskip

We give the principal details of the calculations described in the main text of the Letter. 
\bigskip

I- STOCHASTIC CFT TECHNIQUES 

\noindent
Ia- A simple example $\&$ its connection to random geometry\\
Ib- Operator evolution\\
Ic- Twist operator $\&$ entanglement entropy in stochastic CFT
\medskip

II- ANALYSIS OF STOCHASTIC TRAJECTORIES

\noindent 
IIa- Stochastic processes and reversed stochastic processes\\
IIb- Diffusion in a random flow\\
IIc- Single trajectory behavior\\
IId- Reflection principle for random trajectories in independent fields $\xi^\pm_t$\\
IIe- Jacobian\\
IIf- TT correlation

\bigskip

\section{I- Stochastic CFT techniques}

\subsection{Ia- A simple example $\&$ its connection to random geometry}

The simplest example is that of a massless free Gaussian boson corresponding to a $c=1$ CFT. Our model est then equivalent to that of a free Gaussian field in a random metric specified by the two vector fields $U^\pm_t(x) = \pm\big( v + \dot W^\pm_t(x)\big)$.

It possesses two chiral operators of conformal dimension one, usually called left/right currents (i.e. in the pure CFT in non random environment) but which we choose here to denote $\mathfrak{n}(x)$ and $\bar{\mathfrak{n}}(x)$  and call then {\it densities} for a reason which will be come clear in a short while. 
Quantization is done by imposing that the densities satisfy the Heisenberg-like canonical commutation relations:
\bea
&& [\mathfrak{n}(x),\mathfrak{n}(y)]=+i\delta'(x-y),\\
&& [\bar{\mathfrak{n}}(x),\bar{\mathfrak{n}}(y)]=-i\delta'(x-y),
\eea
while $[\mathfrak{n}(x),\bar{\mathfrak{n}}(y)]=0$. The stress tensor densities are quadratic in the densities:
\be
T(x) = \frac{1}{2} :\mathfrak{n}^2(x):,\quad \bar T(x) = \frac{1}{2} :\bar{\mathfrak{n}}^2(x):.
\ee
This construction is known as the Sugawara's construction and it ensures the stress tensor commutation relations of the Virasoro algebra. By construction, the densities are conformal primary field with dimension $1$ meaning that their commutation relations with the stress tensor components are
\begin{eqnarray}
&& [T[u], \mathfrak{n}(y)]=  +  i\partial_y\big(u(y)\mathfrak{n}(y)\big)\ ,\quad  [\bar T[u], \mathfrak{n}(y)]=0,\\
&& [\bar T[u], \bar{\mathfrak{n}}(y)]=  - i\partial_y\big(u(y)\bar{\mathfrak{n}}(y)\big)\ ,\quad  [T[u], \bar{\mathfrak{n}}(y)]=0,
\end{eqnarray}
with $T[u]:=\int dx\, T(x)\, u(x)$ and $\bar T[u]:=\int dx\, \bar T(x)\, u(x)$ respectively.

To simplify the matter, let first consider the case where the vector fields $\dot W^\pm_t(x)$ are time dependent but not stochastic (the generalisation to stochastic, Brownian driven, vector fields, will be easy). We thus consider Hamiltonian evolution on tis $c=1$ CFT driven by the time dependent Hamiltonian
\be 
H_t = H_0 + \int dx\big( \dot W^+_t(x) T(x) + \dot W^-_t(x) \bar T(x)\big).
\ee

The equations of motion for the primary operators $\mathfrak{n}$ and $\bar{\mathfrak{n}}$ are then
\bea \label{eq:motion-rho}
&& \partial_t\mathfrak{n}(x,t) + \partial_x \big( U^+_t(x) \mathfrak{n}(x,t)\big) =0, \label{eq:motion-rhoA}\\
&& \partial_t\bar{\mathfrak{n}}(x,t) + \partial_x \big( U^-_t(x)\bar{\mathfrak{n}}(x,t)\big) =0, \label{eq:motion-rhoB}
\eea
with $U^\pm_t(x) = \pm\big( v + \dot W^\pm_t(x)\big)$. These are two conservation laws,
\bea 
&&\partial_t\mathfrak{n}(x,t) + \partial_x j(x,t)=0,\\
&& \partial_t\bar{\mathfrak{n}}(x,t) + \partial_x \bar j(x,t)=0,
\eea
with densities $\mathfrak{n}(x,t)$ (resp. $\bar{\mathfrak{n}}(x,t)$) and currents $j(x,t)= U^+_t(x) \mathfrak{n}(x,t)$ (resp. $\bar j(x,t)= U^+_t(x) \bar{\mathfrak{n}}(x,t)$). As in the main text, because these equations are transport equations, they are solved by looking at the backward trajectories associated to the vector fields $U^\pm_t(x)$. The equations for these trajectories are $\dot x^\pm_s= U^\pm_s(x_s)$, as usual. That is
\bea
{\mathfrak{n}}(x,t) = {X^+_t}'(x)\,{\mathfrak{n}}(X^+_t(x)),\\
\bar{\mathfrak{n}}(x,t) = {X^-_t}'(x)\, \bar{\mathfrak{n}}(X^-_t(x)),
\eea
with $X^\pm_t(x)$ the positions at initial time of the trajectories ending at point $x$ at time $t$. They satisfy $\dot X^\pm_t(x)+U_t^\pm(x){X^\pm_t}'(x)=0$.

These equations of motion are equivalent to those of free field $\phi(x,t)$ in a non trivial metric $g_{\mu\nu}$ with action
\bea \label{eq:action0}
S = \frac{1}{2} \int dxdt\, \sqrt{|g|}\, g^{\mu\nu}(\partial_\mu\phi)(\partial_\nu\phi) .
\eea
Indeed, take $g_{\mu\nu}$ be the uni-modular metric, i.e. with $|g|:=|\mathrm{det} g |= 1$, with
\bea \label{eq:metric}
g^{tt}=\frac{2}{U^+-U^-},\ g^{tx}=g^{xt}=\frac{U^++U^-}{U^+-U^-},\ g^{xx}=\frac{2\,U^+U^-}{U^+-U^-},
\eea
with $U^\pm$ the vector fields $U^\pm= \pm( v + \dot W^\pm)$, so that the action reads
\bea \label{eq:action}
 S =  \int \frac{dxdt}{(U^+-U^-)}\Big[ (\partial_t\phi)^2 + (U^++U^-)(\partial_t\phi)(\partial_x\phi) + U^+U^-(\partial_x\phi)^2\Big].
\eea
Equivalently, the metric is proportional to 
\bea \label{eq:Umetric}
ds^2 = g_{\mu \nu} dx^\mu dx^\nu \propto (dx-U^+dt)(dx-U^-dt). 
\eea
The proportionality factor is irrelevant as the action \eqref{eq:action0} is conformally invariant. But, if one wishes to normalize the metric to be uni-modular with $\mathrm{det}g=-1$ (as for a 2d metric with Lorentzian signature), then $ds^2=\frac{2}{U^+-U^-}(dx-U^+dt)(dx-U^-dt)$. The two vector fields $U^\pm$ specify locally the light-cone since $ds^2=0$ for $dx=U^+dt$ or $dx=U^-dt$. These last two equations are the local flow equations associated to the vector fields $U^\pm$, so that the trajectories of the $U^\pm$-flows are actually the null geodesics of this geometry.

The Euler-Lagrange equations of motion, obtained by extremising the action \eqref{eq:action}, are then
\bea \label{eq:E-L}
\partial_t\big[ (\nabla_+\phi) +(\nabla_-\phi) \big] + \partial_x\big[ U^+(\nabla_+\phi)+U^-(\nabla_-\phi)\big]=0,
\eea
with
\bea
\nabla_\pm := \frac{1}{U^+-U^-}\big[ \partial_t + U^\mp \partial_x\big] .
\eea
Eq.\eqref{eq:E-L} is a local conservation for a density $(\nabla_+\phi) +(\nabla_-\phi)$ and a current $U^+(\nabla_+\phi)+U^-(\nabla_-\phi)$. As usual, one has also the conservation law for the topological current $\epsilon^{\mu\nu}\partial_\nu\phi$, which reads
\bea \label{eq:J-top}
\partial_t\big[ (\nabla_+\phi) - (\nabla_-\phi) \big] + \partial_x\big[ U^+(\nabla_+\phi)-U^-(\nabla_-\phi)\big]=0,
\eea
Eq.\eqref{eq:J-top} is tautologically (or topologically) fulfilled because $(\nabla_+\phi) - (\nabla_-\phi) =-\partial_x\phi$ and $U^+(\nabla_+\phi)-U^-(\nabla_-\phi)=\partial_t\phi$. Hence Eqs.(\ref{eq:E-L},\ref{eq:J-top}) are the sum and the difference of the equations of motion (\ref{eq:motion-rhoA},\ref{eq:motion-rhoB}) for ${\mathfrak{n}}$ and $\bar{\mathfrak{n}}$, with
\bea
\mathfrak{n} = \nabla_+\phi\ ,\quad \bar{\mathfrak{n}} = \nabla_-\phi .
\eea

Notice that any (generic) Lorentzian uni-modular metric can be parametrized in terms of two vector fields as in \eqref{eq:metric}. Since any 2d metric (on a surface of genus zero) is conformally equivalent to an uni-modular metric, any 2d massless free field theory can be viewed as coding for transport along a non trivial vector fields.

As mentioned in the main text and explained below (in the case of any chiral operators), this result holds true also if the vector fields $\dot W_t^\pm(x)$ are stochastic, driven by Brownian motion, say of the form $\dot W^\pm_t(x)= \xi_t^\pm(x) + \sqrt{D_0}\dot B^\pm_t$. However, the presence of the Brownian motion $B^\pm_t$ implies that the mean operators (mean with respect to those Brownian motions) satisfy a diffusive equation.

To simplify the matter even further (and to be able to do explicit simple computation in the stochastic case), let us now consider the case $\xi^\pm \equiv 0$, so that $\dot W^\pm_t(x)= \sqrt{D_0}\dot B^\pm_t$ and $U^\pm_t(x)= \pm v \pm  \sqrt{D_0}\dot B^\pm_t$, with $B^\pm_s$ two independent normalized Brownian motions. The trajectories $x^\pm_s$ are then described by the explicit equations:
\be
x_s^\pm = x_0 \pm v s \pm \sqrt{D} B^\pm_s,
\ee
so that
\be
X^\pm_t(x)= x \mp v t \mp \sqrt{D_0} B^\pm_t.
\ee
In this simple case, ${X_t^\pm}'(x) =1$, independently of $x$, and there is no Jacobian factors in the solutions of the equations of motion so that $\mathfrak{n}(x,t) = \mathfrak{n}(X^+_t(x))$ and $\bar{\mathfrak{n}}(x,t) = \bar{\mathfrak{n}}(X^-_t(x))$

We can compute (exactly) the correlation functions of product the densities $\mathfrak{n}$ and $\bar{\mathfrak{n}}$. For instance, the two-point correlation of the density $\mathfrak{n}$, in the Calabrese-Cardy state, reads (how to compute expectation with the Cardy-Calabrese state in explained in the following Appendix Ic)
\bea
&&\langle \Psi_0| \mathfrak{n}(x_2,t_2)\mathfrak{n}(x_1,t_1)|\Psi_0\rangle = \langle \Psi_0| \mathfrak{n}(X_{t_2}(x_2))\mathfrak{n}(X_{t_1}(x_1))|\Psi_0\rangle\\
&& \hskip 4.0 truecm = \frac{\mathrm{const.} }{(\cosh( \frac{\pi}{4 v \tau_0} (x_1-x_2 -v(t_1-t_2) - \sqrt{D_0}(B^+_{t_2}-B^+_{t_1})))^2}
\eea
Suppose that $t_2>t_1$. Then, $B^+_{t_2}-B^+_{t_1}\equiv B^+_{t_2-t_1}$  in law, so that an exact integral representation of moments of this correlation can be written.

Notice that, using again the tools discussed in the Appendix Ic
about the Calabrese-Cardy construction, a similar formula can be written for the correlation of the twist operators (in the limit $x_0\to-\infty$) and hence for the exponential moments of the entanglement entropy. Indeed, since $X_t^\pm(x) = x \mp v t \mp \sqrt{D} B^\pm_t$ for $\xi=0$, the Jacobian $J_t(x_0,x) =1$ and
 \bea
 && G_t(x_0,x) \simeq  
 \frac{A_\phi^2 (\frac{\pi}{4 v \tau_0})^{4 \Delta_n} }{[ \cosh( \frac{\pi (2 v t +\sqrt{D} (B^-_t - B^+_t))}{4 v \tau_0} ) ]^{4 \Delta_n}} \nonumber
 \eea 
and 
\bea
\langle e^{-q S_n } \rangle_B \simeq \tau_0^{-4 q \delta_n} 
\int dy^+ dy^- 
\frac{P^+_{t,t_0}(x,y^+) P^-_{t,t_0}(x,y^-) }{\cosh(\frac{\pi (y^+-y^-)}{4 v \tau_0} )^{2 q \delta_n}}
\eea
with $P^\pm_{t,t_0}(x,y^\pm)$ the distribution of Brownian trajectories as defined in the text. The analysis of the $q$-dependence of the above expectation is then parallel to that discussed in the main text.

\subsection{Ib- Operator evolution}

Here we show how the evolution equations for chiral (resp. anti-chiral) operators in stochastic CFTs are solved by eq.\eqref{evolution}. For simplicity we set $v=1$ in this Section.

Recall that the two (chiral / anti-chiral) components $T$ and $\bar{T}$ of the stress tensor in CFT satisfy the Virasoro commutation relations
\begin{eqnarray}
&& [T(x),T(y)]  = -i\Big(  2\delta'(x-y) T(y) - \delta(x-y) T'(y)  - \frac{c}{24\pi}\, \delta{'''}(x-y)\Big),  \\
&& [ \bar{T}(x), \bar{T}(y) ]   = +i\Big( 2\delta'(x-y) \bar{T}(y) - \delta(x-y) \bar{T}'(y)  - \frac{c}{24\pi}\, \delta{'''}(x-y)\Big),
 \end{eqnarray}
with $c$ the central charge and where the prime symbol denotes derivative with respect to the space variable. Alternatively (and for later convenience), we defined $T_\pm[u]:=\int dx\, T_\pm(x)\, u(x)$, for any vector field $u(x)$, so that the commutation relations read
\[ [T[u], T(y)]= + i\Big( 2u'(y)T(y) + u(y)T'(y) - \frac{c}{24\pi}\, u{'''}(y) \Big) ,\]
and similarly (up to a global sign) for $\bar T$. Similarly one can write the commutator $[T[u_1],T[u_2]]$ for any pair of vector fields $u_1$ and $u_2$. 
The normalisation is such that, defining $\mathcal{L}_k:=\int dx e^{-ikx} T(x)$, their commutation relations are
\[ \big[\mathcal{L}_k,\mathcal{L}_p\big] = (k-p) \mathcal{L}_{k+p} +\frac{c}{12}\, k^3\, \delta(k+p).\]

Recall also that an operator $\varphi(x)$ is a chiral primary operator of dimension $\Delta$ if, by definition, it satisfies the following commutation relations with the stress tensor:
\[ [T[u],\varphi(x) ] = i\Big( \Delta\, u'(x)\varphi(x) + u(x)\varphi'(x) \Big),\quad [\bar T[u],\varphi(x) ] =0.\]
A similar definition holds for the anti-chiral primary operators, with the role of $T$ and $\bar T$ exchanged.

Recall that, in the stochastic CFT model we consider, the time evolution of operators is defined as $O\to O(t):= {\cal U}_t^\dag\, O\, {\cal U}_t$, where the unitary ${\cal U}_t$ evolves according to ${\cal U}_{t+dt} {\cal U}^\dagger_t = e^{- i dH_t}$, with Hamiltonian increment $dH_t = H_0 dt +  \int dx \, (dW^+_t(x) T(x) + dW^-_t(x) \bar T(x))$ where $dW^\pm_t(x) = \xi^\pm_t(x) dt + \sqrt{D} dB^\pm_t$  as in the main text, eq.\eqref{dHt}. By expanding up to order $dt$, this leads to the evolution equation for $O(t)$:
\[ dO(t) = +i [dH_t,O](t) -\frac{1}{2} [dH_t, [dH_t,O]](t) .\]
Recall that $O(t):= {\cal U}_t^\dag\, O\, {\cal U}_t$, so that here $[dH_t,O](t)= {\cal U}_t^\dag\, [dH_t,O]\, {\cal U}_t$ with $[dH_t,O]:=dH_t\,O-O\, dH_t$ the commutator of the two operators $dH_t$ and $O$, and similarly for the double commutator $[dH_t, [dH_t,O]](t)$.
In particular, for chiral operators (i.e. for $O\to \varphi(x)$), we have:
\be
d\varphi(x,t) = +i [T[dU^+_t],\varphi(x)](t) -\frac{1}{2} [T[dU^+_t], [T[dU^+_t],\varphi(x)]](t), 
\ee
with vector field increment $dU^+_t(x)=vdt+dW^+_t(x)$.

Let us now write these equations in a more compact way. We define the differential operators $\mathcal{D}_\Delta[u]$ by their action on functions $f(x)$:
\[ \mathcal{D}_\Delta[u]\cdot f(x) := \big( \Delta\, {u}'(x)+ u(x) \partial_x\big) f(x) .\]
The definition of $\mathcal{D}_\Delta$ is made such that the commutation relations of chiral operators with the stress tensor simplifies~:
\be
[T[u],\varphi(x)] = i\,\mathcal{D}_\Delta[u]\cdot\varphi(x).
\ee
Hence, the equation of chiral primary operator may be written as
\be \label{eq:motion}
d\varphi(x,t) + \mathcal{D}_\Delta[dU^+_t]\cdot\varphi(x,t) -\frac{1}{2}(\mathcal{D}_\Delta[dU^+_t])^2\cdot\varphi(x,t) =0 .
\ee
Using $dU^+_t(x)d{U^+_t}'(x)=0$ and $d{U^+_t}'(x)^2+dU^+_t(x)d{U^+_t}''(x)=0$ by translation invariance of the statistics of the velocity fields, one checks that $(\mathcal{D}_\Delta[dU^+_t])^2\cdot f(x) = \Big( \Delta(\Delta-1) d{U^+_t}'(x)^2+ dU^+_t(x)^2 \partial_x^2\Big) f(x)$.

We can then prove that the solution of equation \eqref{eq:motion} is 
\be \label{eq:solution}
\varphi(x,t) = \big[ X_t'(x)\big]^\Delta\, \varphi(X_t(x)),
\ee
where $X_t(x)$ is the initial position of the trajectories with vector field increment $dU^+_t(x)$ which will be at time $t$ at position $x$. That is:  $t\to X_t(x)$ is the backward trajectory. 

Indeed, as explained in the main text, by definition this initial position satisfies
\be
X_{t+dt}(x + dU^+_t(x))= X_t(x)
\ee
However, one has to be careful when dealing with the last equation because  $X_{t+dt}(\cdot)$ and $dU^+_t(x)$ are not independent. One has first to invert this relation (i.e. the relation between $x$ and $y=x+dU^+_t(x)$) and write $X_{t+dt}(x)$ in terms of $X_t(\cdot)$. The result is $X_{t+dt}(x)=X_t(x-dU^+_t(x)+\cdots)$, using $dU^+_t(x)d{U^+_t}'(x)=0$. Then, one may Taylor expand this last relation to second order using It\^o rules to prove that $X_t(x)$ is the solution of the following stochastic diffusive transport equation:
\be \label{eq:dX}
dX_t(x) + dU^+_t(x)\, X_t'(x) -\frac{1}{2} d{U_t^+}(x)^2\, X_t''(x)= 0, 
\ee
with initial condition $X_{t=0}(x)=x$. 
Let us now look at the equation of motion \eqref{eq:motion} but first for $\Delta=0$ to simplify the discussion. It reads
\bea \label{eq:motion0}
d\varphi(x,t) + dU^+_t(x)\, \partial_x\varphi(x,t) -\frac{1}{2} d{U_t^+}(x)^2\, \partial_x^2\varphi(x,t)= 0.
\eea
To prove the claim \eqref{eq:solution} for $\Delta=0$, we have to prove that $\varphi(x,t):=\varphi_0(X_t(x))$ is a solution of the above equation for any $\varphi_0$. Let us compute the It\^o and space derivatives of this field $\varphi(x,t)$. By It\^o rules, we have (for $\Delta=0$):
\begin{eqnarray}
d\varphi(x,t) &=& dX_t(x)\, \varphi_0'(X_t(x)) + \frac{1}{2} dX_t(x)^2\, \varphi_0''(X_t(x)),\\
\partial_x\varphi(x,t) &=& X_t'(x)\, \varphi_0'(X_t(x)),\\
\partial_x^2\varphi(x,t) &=& X_t''(x)\, \varphi_0'(X_t(x))+ X_t'(x)^2\, \varphi_0''(X_t(x)).
\end{eqnarray}
Thus, equation \eqref{eq:motion0} is going to be fulfilled for any $\varphi_0$ if the following two equations are satisfied:
\begin{eqnarray}
&& dX_t(x) + dU^+_t(x)\, X_t'(x) -\frac{1}{2} d{U_t^+}(x)^2\, X_t''(x)= 0,\\
&& dX_t(x)^2 - dU^+_t(x)^2\, X_t'(x)^2 = 0.
\end{eqnarray}
The first is simply the characteristic equation \eqref{eq:dX} for the backward trajectories, that we just proved, and the second is a consequence of the first (simply by taking the square). Thus, $\varphi(x,t)=\varphi_0(X_t(x))$ is indeed a solution of the stochastic equation of motion for $\Delta=0$.

The case $\Delta\not= 0$ is done similarly. We aim at proving that $\varphi(x,t)=[Y_t(x)]^\Delta\, \varphi_0(X_t(x))$, with $Y_t(x):= X_t'(x)$ is a solution of the equations of motion \eqref{eq:motion}. Assuming translation invariance of the covariance of the velocity fields, the latter read (to simplify the notation we set $U_t^+(x) \to U_t(x)$, without the upper-script $+$)
\bea \label{eq:phiDelta}
d\varphi(x,t) + dU_t(x)\, \partial_x\varphi(x,t) + \Delta dU_t'(x)\, \varphi(x,t)
 = \frac{1}{2} d{U_t}(x)^2\, \partial_x^2\varphi(x,t) +  \frac{1}{2}\Delta(\Delta-1) d{U_t'}(x)^2\, \varphi(x,t).
\eea
Let us first compute the It\^o derivative $d\varphi(x,t)$ (without forgetting the crossed terms),
\begin{eqnarray}
d\varphi(x,t)&=& [Y_t(x)]^\Delta\, \Big[ dX_t(x)\varphi_0'(X_t(x))+ \frac{1}{2} dX_t(x)^2\varphi_0''(X_t(x)) \\
&& + \Delta\, Y_t(x)^{-1}dY_t(x)dX_t(x)\,\varphi_0'(X_t(x)) + \Big(  \Delta\, Y_t(x)^{-1}dY_t(x)+ \frac{1}{2}\Delta(\Delta-1)\, Y_t(x)^{-2}dY_t(x)^2\Big)\varphi_0(X_t(x))\Big] . \nonumber
\end{eqnarray}
Let us now compute the space derivatives,
\begin{eqnarray}
\partial_x\varphi(x,t)&=& [Y_t(x)]^\Delta\, \Big[ X_t'(x)\varphi_0'(X_t(x)) + \Delta\, Y_t(x)^{-1}Y_t'(x)\,\varphi_0(X_t(x))\Big],\\
\partial_x^2\varphi(x,t)&=& [Y_t(x)]^\Delta\, \Big[ X_t'(x)^2\varphi_0''(X_t(x)) + \Big( X_t''(x) + 2\Delta\, Y_t(x)^{-1}Y_t'(x)X_t'(x)\Big) \varphi_0'(X_t(x))\\
&& + \Big( \Delta\, Y_t(x)^{-1}Y_t''(x) + \Delta(\Delta-1)\, Y_t(x)^{-2}Y_t'(x)^2\Big)\varphi_0(X_t(x)) \Big].\nonumber 
\end{eqnarray}
Thus, comparing the last three equations and using the characteristic equation \eqref{eq:dX} for $X_t(x)$, namely $dX_t(x) + dU_t(x)\, X_t'(x) =\frac{1}{2} d{U_t}(x)^2\, X_t''(x)$, the equations of motion \eqref{eq:phiDelta} is solved by $\varphi(x,t)=[X_t'(x)]^\Delta\, \varphi_0(X_t(x))$ iff
\begin{eqnarray}
dY_t(x) + dU_t(x) Y_t'(x) + dU_t'(x) Y_t(x) &=& \frac{1}{2} dU_t(x)^2\,Y_t''(x) \\
dY_t(x) dX_t(x) &=& dU_t(x)^2\, Y_t'(x)X_t'(x) ,\\
dY_t(x)^2 &=&  dU_t(x)^2\, Y_t'(x)^2 + dU_t'(x)^2\, Y_t(x)^2 ,
\end{eqnarray}
with $Y_t(x)=X_t'(x)$, the space derivative of $X_t(x)$. The first equation is a consequence of \eqref{eq:dX} (by taking the space derivative of \eqref{eq:dX}), the second is a consequence of the first and \eqref{eq:dX} by multiplying them, and the third is also a consequence of the first by squaring it (taking into account in the three equations that $dU_t(x)dU_t'(x)=0$ by translation invariance). Everywhere, we were allowed to
neglect terms $O(dt^{3/2})$ and higher.
Thus, we proved that $\varphi(x,t)=[Y_t(x)]^\Delta\, \varphi_0(X_t(x)$ is solution of the stochastic equations of motion with the initial condition $\varphi(x,t=0)= \varphi_0(x)$ (since $X_{t=0}(x)=x$, and hence $X_{t=0}'(x)=1$).

Similar results apply the anti-chiral operators with $dU^+_t$ replaced by $dU^-_t(x):=-(vdt + dW^-_t(x))$.

Similar results also apply the stress-tensor components. One may prove that the equations of motion of the stress-tensor are solved by
\be \label{TT} 
T(x,t) = [X_t'(x)]^2\, T(X_t(x)) - \frac{c}{24\pi} \, (S\cdot X_t)(x) ,
\ee
where $(S\cdot f)$ denotes the Schwarzian derivative of the function $f$: $(S\cdot f)(x) = \frac{ f'''(x)}{f'(x)}-\frac{3}{2} \big(\frac{f''(x)}{f'(x)}\big)^2$. This result follows from the stochastic equation satisfies by $X_t(x)$ and from the cocycle relation fulfilled by the Schwarzian derivative: $(S\cdot (g\circ f))(x) = f'(x)^2\, (S\cdot g)(f(x)) + (S\cdot f)(x) $. The proof of this statement is a bit more delicate as for \eqref{eq:solution}, as it relies on this cocycle relation but, once one knows that this relation together with \eqref{eq:dX} are the key ingredients, the proof of \eqref{TT} is a consequence of It\^o calculus.

All these results are direct consequences of the fact that the stress-tensor is the generator of the diffeomorphism.

\subsection{Ic- Twist operators \& entanglement entropy in stochastic CFT}
\label{app:Ic} 

Here we show how to compute the entanglement entropy for a semi-infinite line in stochastic CFT.

Recall from \cite{Cardy1} that, in quantum field theory and for a system initially prepared in a pure state $|\Psi_0\rangle$, the entanglement entropy, or the Reniy entropy $S_n(x_0,x|t)$, of an interval $[x_0,x]\subset \mathbb{R}$ may be represented as 
\be \label{eq:S-cft}
e^{(1-n)S_n(x_0,x|t)} = \langle \Psi_0|\hat \Phi_n(x_0,t)\Phi_n(x,t)|\Psi_0\rangle
\ee
where $\Phi(x)$ and $\hat \Phi(x_0)$ are the so-called conjugated twist operators. Here $\Phi(x,t)$ denote the time evolved twist operator $\Phi(x)$ at time $t$. 

We apply this formula in (stochastic) CFT with $|\Psi_0\rangle$ a `gapped' state with small coherence length.

Recall the representation of gapped states with small coherence length introduced by Calabrese and Cardy \cite{CalabreseCardy1,CalabreseCardy2}. They formally write this state as $|\Psi_0\rangle \propto e^{-\tau_0\, H_0}|B\rangle\rangle$ with $H_0$ the CFT hamiltonian and $|B\rangle\rangle$ a (unormalisable) boundary state. Here, $\tau_0$ is parametrising the coherence length, which vanishes as $\tau_0\to 0$. 

This representation means the following rules:\\
(i) The expectation of any product of localised primary operators 
\be
\langle \Psi_0| O_1(x_1)\cdots O_M(x_M)|\Psi_0\rangle
\ee
is defined by analytic continuation from the Euclidean CFT correlation functions on the infinite strip of width $2\tau_0$, embedded into the complex plane, with main axis along the real axis and its two boundaries parallel to the real axis and crossing the imaginary axis at points $\pm i \tau_0$. The operators $O_j(x_j)$ are positioned at points $x_j$ along the real axis inside this strip. The boundary conditions imposed on the two strip boundaries are those encoded in the state $|B\rangle\rangle$.\\
(ii) By conformal invariance, the strip CFT expectation values are mapped into CFT expectation values in the upper half plane (UHP). If $w=x+i\tau$ is the complex parameter in the strip ($x\in \mathbb{R}, \tau\in[-\tau_0,\tau_0]$), the conformal mapping to the UHP is 
\[ w\to z= i \exp({\frac{\pi}{2\tau_0} w}),\]
so that the points on the lower boundary $\tau=-\tau_0$ are mapped to the positive real axis, and those on the upper boundary $\tau=+\tau_0$ are mapped on the negative real axis. The covariance property of the CFT expectation values depends on the scaling dimensions of the operators.\\
(iii) Correlation functions in boundary CFTs, and hence in the UHP, are computed using the method of images. This amounts to decompose the operators $O_j$ in its chiral and anti-chiral components, $O(z)=\bar \varphi_O(z^-)\, \varphi_O(z^+)$, where $z^+$ (resp. $z^-$) are the images of $z$ in the upper half plane (resp. lower half plane), and to write the boundary CFT correlation functions in terms of the conformal blocks
\be
\langle  \bar \varphi_{O_1}(z_1^-)\, \varphi_{O_1}(z_1^+) \cdots \bar \varphi_{O_M}(z_M^-)\, \varphi_{O_M}(z_M^+) \rangle_{UHP_B},
\ee
where the (analytical) structure of this conformal block is specified by the boundary condition $|B\rangle\rangle$. 

Let us apply these rules for the expectation \eqref{eq:S-cft}. Decomposing the twist operators in their chiral/anti-chiral components, $\varphi_n$ and $\bar \varphi_n$, yields
\be \label{eq:cft-corr}
\langle \Psi_0|\hat \Phi_n(x_0,t)\Phi_n(x,t)|\Psi_0\rangle = (\frac{\pi}{2\tau_0})^{4\Delta_n}\cdot (z_{x_0,t}^-z_{x_0,t}^+z_{x,t}^-z_{x,t}^+)^{\Delta_n}\cdot
\langle  \hat{\bar \varphi}_{n}(z_{x_0,t}^-)\, \hat{\varphi}_n(z_{x_0,t}^+) \, \bar \varphi_n(z_{x,t}^-)\, \varphi_n(z_{x,t}^+) \rangle_{UHP_B}
\ee
where $z_{x,t}^+$ (resp. $z_{x,t}^-$) is the image in the upper half plane (resp. lower half plane) of the positions of the chiral (resp. anti-chiral) components of the time evolved twist operator $\Phi_n(x,t)$.  And similarly for the conjugated twist operator $\hat \Phi_n(x_0,t)$. The first two factors arise from the Jacobians of the conformal transformation $w\to z$. (Notice that $z'(w)= \frac{\pi}{2\tau_0}\, z$.)

In absence of stochastic disorder (i.e. for pure CFT), these images would be $z_{x,t}^+= ie^{\frac{\pi}{2\tau_0}(x-t)}$ and $z_{x,t}^-=-ie^{\frac{\pi}{2\tau_0}(x+t)}$. In presence of stochastic disorder, due to the covariance property \eqref{evolution} of the chiral/anti-chiral operators, they are given by
\be
z_{x,t}^+=i\, e^{\frac{\pi}{2\tau_0}\, X^+_t(x)}, \quad z_{x,t}^-= - i\, e^{\frac{\pi}{2\tau_0}\, X^-_t(x)},
\ee
where $X_t^\pm(x)$ are the backward trajectories introduced in the main text.

By conformal invariance, the conformal block \eqref{eq:cft-corr} can be written as
\be
\langle \Psi_0|\hat \Phi_n(x_0,t)\Phi_n(x,t)|\Psi_0\rangle = (\frac{\pi}{2\tau_0})^{4\Delta_n}\cdot \left(\frac{z_{x_0,t}^-z_{x_0,t}^+z_{x,t}^-z_{x,t}^+}{(z_{x_0,t}^+-z_{x,t}^+)^2(z_{x_0,t}^--z_{x,t}^-)^2}\right)^{\Delta_n}\cdot
K_B(\eta)
\ee
where $\eta$ is the harmonic ratio
\be
\eta = \frac{(z_{x_0,t}^+-z_{x_0,t}^-)(z_{x,t}^+-z_{x,t}^-)}{(z_{x_0,t}^+-z_{x,t}^-)(z_{x,t}^+-z_{x_0,t}^-)}.
\ee

In general, the conformal block $K_B(\eta)$ is difficult to compute since one needs extra information to determine it. However, it simplifies in the limit $\eta\ll 1$ as one can then use operator product expansions (OPE). In the present context, the condition $\eta\ll 1$ is going to be realised when $x_0\to -\infty$ because then $X_t(x_0)\to -\infty$ almost surely (a.s.) so that both $z_{x_0,t}^+$ and $z_{x_0,t}^-$ approach $0$ along the imaginary axis, one from above, the other from below. As a consequence, when $x_0\to -\infty$, one can use the operator product expansion, 
\be \label{eq:ope-b}
\hat{\bar \varphi}_{n}(z_{x_0,t}^-)\, \hat{\varphi}_n(z_{x_0,t}^+) \simeq \frac{A_\phi}{(z_{x_0,t}^--z_{x_0,t}^+)^{2\Delta_n-\Delta_b}}\, \Psi_b(0) + \cdots,
\ee
to reduce the computation of the conformal block \eqref{eq:cft-corr} to that of a three point function with insertion of the boundary operator $\Psi_b(0)$. Namely, to the three point function $\langle \Psi_b(0)\, \bar \varphi_n(z_{x,t}^-)\, \varphi_n(z_{x,t}^+) \rangle_{UHP_B}$, which is exactly known by conformal invariance. The output can be summarised by
\be
K_B(\eta) \simeq A_\phi^2\, \frac{1}{\eta^{2\Delta_n-\Delta_b}} + \cdots,
\ee
for $\eta\ll 1$, with $\Delta_n$ the scaling dimension of $\varphi_n$ and $\Delta_b$ that of $\Psi_b$.

Hence, for $x_0\to -\infty$, we have
\be
\langle \Psi_0|\hat \Phi_n(x_0,t)\Phi_n(x,t)|\Psi_0\rangle \simeq (\frac{\pi}{2\tau_0})^{4\Delta_n}\cdot A_\phi^2\cdot \frac{ (z_{x_0,t}^-z_{x_0,t}^+)^{\Delta_n}\, (z_{x,t}^-z_{x,t}^+)^{\Delta_n-\Delta_b}}{[(z_{x_0,t}^+-z_{x_0,t}^-)(z_{x,t}^+-z_{x,t}^-)]^{2\Delta_n-\Delta_b}}.
\ee
Using $z_{x,t}^\pm=\pm i\, e^{\frac{\pi}{2\tau_0}\, X^\pm_t(x)}$, this can be rewritten as
\be
\langle \Psi_0|\hat \Phi_n(x_0,t)\Phi_n(x,t)|\Psi_0\rangle \simeq (\frac{\pi}{2\tau_0})^{4\Delta_n}\cdot A_\phi^2\cdot 
\frac{ e^{\frac{\pi}{4\tau_0}(X_t^+(x_0)+X_t^-(x_0)-X_t^+(x)-X_t^-(x))\Delta_b} }{\Big(4\cosh[\frac{\pi}{4\tau_0}(X_t^+(x_0)-X_t^-(x_0))]\,\cosh[\frac{\pi}{4\tau_0}(X_t^+(x)-X_t^-(x))]\Big)^{2\Delta_n-\Delta_b}}
\ee
In ref.\cite{CalabreseCardy1,CalabreseCardy2}, it was proved that $\Delta_b=0$, so that 
\be
\langle \Psi_0|\hat \Phi_n(x_0,t)\Phi_n(x,t)|\Psi_0\rangle \simeq
\frac{ (\frac{\pi}{2\tau_0})^{4\Delta_n}\cdot A_\phi^2 }{\Big(4\cosh[\frac{\pi}{4\tau_0}(X_t^+(x_0)-X_t^-(x_0))]\,\cosh[\frac{\pi}{4\tau_0}(X_t^+(x)-X_t^-(x))]\Big)^{2\Delta_n}}
\ee
as claimed in the main text.

\section{II- Analysis of stochastic trajectories}

\subsection{IIa- Stochastic processes and reversed stochastic processes}

Here we gather a few (standard) information on stochastic processes (driven by Brownian motions) and their time reversed.
Let $u \to x_u$ be a stochastic process defined by the stochastic equation
\be \label{eq:sde}
 d x_u = U_u(x_u)\, du + \sqrt{D}dB_u,
 \ee
with $B_u$ a normalised Brownian motion, $dB_u^2=du$. To simplify the description, we here assume the drift $U_u(x)$ to be smooth enough, both in time and space. If not too strong, non-smoothness can (in some cases) be controlled, say as in turbulent transport \cite{BernardGawedzkiKupiainen}, but requires an alternative construction, as in \cite{LeJanRaimond}, if too strong, because then the increments along a trajectory are highly erratic.

Given any realisation of the Brownian motion, the forward process consists is fixing the initial condition $x_{u=t_0}=x_0$ and solving the SDE \eqref{eq:sde} forward in time. One then looks at the distribution of the position $x_t$ at a later time $u=t$
\[ \mathbb{P}_\uparrow(x,t|x_0,t_0) = \mathbb{E}[\delta(x_t-x)] .\]
It is well know that it satisfies the Fokker-Planck equation and its dual $(t_0<t)$:
\bea
 \partial_t \mathbb{P}_\uparrow(x,t|x_0,t_0)  && = \Big( \frac{D}{2}\partial_x^2 - \partial_x\, U_t(x)\Big) \mathbb{P}_\uparrow(x,t|x_0,t_0),\\
\partial_{t_0} \mathbb{P}_\uparrow(x,t|x_0,t_0)  &&  = - \Big( \frac{D}{2}\partial_{x_0}^2 +U_{t_0}(x_0) \partial_{x_0}\, \Big) \mathbb{P}_\uparrow(x,t|x_0,t_0),
\eea
with initial condition $\mathbb{P}_\uparrow(x,t_0|x_0,t_0) = \delta(x-x_0)$.

The reversed process is defined (for a fixed $t$) as $y_s := x_{t-s}$ (with $s\in[0,t-t_0]$) so that $y_{s=0}=x_t$ and $y_{s=t-t_0}=x_{t_0}$. For any realisation of the Brownian motion, it consists in fixing the final position of $x_\cdot$ (that is, the initial position of $y_s$), say $y_{s=0}=y_0$, and integrating the SDE \eqref{eq:sde} backward in time. One then look at the distribution of $y_t$ (that is, that of the initial position $x_{t_0}$),
\[ \mathbb{P}_\downarrow(y,s|y_0,0) = \mathbb{E}[\delta(y_s-y)] .\]
The claim is that the process $s\to y_s$ satisfies the SDE
\be \label{eq:sde-2}
 d y_s = \tilde U_s(y_s)\, ds + \sqrt{D}d\tilde B_s,
 \ee
 with $\tilde U$ the time-reversed field, $\tilde U_s(x) := - U_{t-s}(x) $ and $\tilde B_s$ the time reversed Brownian motion (which is also a normalized Brownian motion). As a consequence, $ \mathbb{P}_\downarrow(y,s|y_0,0)$ satisfies the reversed Fokker-Planck equation:
\be \label{eq:dual-eq}
 \partial_s \mathbb{P}_\downarrow(y,s|y_0,0) = \Big( \frac{D}{2}\partial_y^2 - \partial_y\, \tilde U_s(y)\Big) \mathbb{P}_\downarrow(y,s|y_0,0),
\ee
with the reversed field
\be \label{eq:reversed}
\tilde U_s(x) := - U_{t-s}(x) .
\ee

{\bf Connection with the text}.
The process $u \to x_u$ in Eq. \eqref{langevin} in the text corresponds to the 
forward process $u \to x_u$ in Eq. \eqref{eq:sde}, with $U_u(x)= \pm (v + \xi_u^\pm(x))$.
The process $s \to X^\pm_{t,t_0=t-s}(x)$ defined in the text, corresponds to the reversed process \eqref{eq:sde-2} with $y_s=X^\pm_{t,t-s}(x)$
and $\tilde U_s(y)=- U_{t-s}(y)=\mp (v + \xi_{t-s}^\pm(y))$. It thus obeys the stochastic
equation
\be \label{Xtt} 
d X^\pm_{t,t-s}(x) = \mp (v + \xi^\pm_{t-s}(X^\pm_{t,t-s}(x)) ) ds +
\sqrt{D} d\tilde B^\pm_{s} 
\ee
Its associated Fokker-Planck equation, Eq. \eqref{FP} in the text, corresponds to Eq. \eqref{eq:dual-eq} 
with $\tilde U_s(y)=\mp (v+ \xi_{t-s}^\pm(y))$ and 
$P_{t,t_0}(x,y) = \mathbb{P}_\downarrow(y,s|x,0)$ with $s=t-t_0$ (hence $dt_0=-ds$).
\medskip

We now give three ways to approach this problem: via stochastic calculus, via time discretisation, via path integral. 

Let us start with stochastic calculus approach. Let $y_s=x_{t-s}$ and look at $dy_s:= y_{s+ds}-y_s$. We have
\begin{eqnarray*}
dy_s  && = x_{t-s-ds}-x_{t-s} = -( x_{t-s-ds+ds}-x_{t-s-ds} ) \\
&& = - U_{t-s-ds}(x_{t-s-ds})\, ds - dB_{t-s-ds}\\
&&=  - U_{t-s}(y_s)\, ds + d\tilde B_s,
\end{eqnarray*}
where we set $\tilde B_s=-B_{t-s}$, we used the smoothness of the function $U_u(x)$ and neglect terms $O(ds^{3/2})$. This proves that the process $y_s$ satisfy the SDE \eqref{eq:sde-2}, and hence, his transition kernel satisfies \eqref{eq:dual-eq}.

Let us now use a time discrete realisation. So we imagine discretising the time interval $[t_0,t]$ is small piece of time step $\epsilon$, with $t_0<t_1<\cdots<t_n=t$. The data of a Brownian sample is replaced by the data of normalised gaussian variables $(b_1,b_2,\cdots,b_{n})$, with $n\epsilon = |t-t_0|$. The SDE \eqref{eq:sde} is replaced by the random updatings of the position via the recursion relations
\be \label{eq:sde-discret}
x_{k+1} = x_k + \epsilon\, U_k(x_k) + \sqrt{D\epsilon}\, b_{k+1} ,
\ee
where, with a slight abuse of notation, we set $U_k(\cdot)\equiv U_{t_k}(\cdot)$ in this equation.

Looking at the forward trajectory consists in starting with $x_0$,  iterating the relations \eqref{eq:sde-discret} to recursively get the position $x_k$, and looking for the distribution of the final position $x_n$.

The backward trajectory consists in starting at the final position $x_f$ and iterating backward the relations \eqref{eq:sde-discret}. That is, we have to invert this relation, expressing $x_k$ in terms of $x_{k+1}$. We get
\be
x_k = x_{k+1} - \epsilon\, U_k(x_{k+1}) - \sqrt{D\epsilon}\, b_{k+1} + O(\epsilon^{3/2}).
\ee
Hence, setting
\be 
y_p=x_{n-p},\quad \tilde U_p(\cdot) = - U_{n-p+1}(\cdot),\quad \tilde b_p=-b_{n-p+1},
\ee
we can write the backward discret updatings (equation above) as
\be \label{eq:discrete-reversed}
y_{p+1} = y_p + \epsilon\, \tilde U_p(y_p) + \sqrt{D\epsilon}\, \tilde b_{p+1} .
\ee
This is the discrete version of \eqref{eq:sde-2}.

The one fact about the process $k\to y_k$ that we have to check is that it indeed describes the reversed trajectory. That is, we have to answer the question : when iterating the discrete forward process starting from position $x_0$, and then the discrete backward process starting from position $x_n$, does the final position coincide the initial position?
Let us first check this fact when there is only one step. The forward process is $x_0=x_*$ and $x_1=x_*+\epsilon\, U_1(x_*) + \sqrt{D\epsilon}\, b$, with $b$ a normalised Gaussian variable. The backward process is $y_0=y_*$ and $y_1= y_* - \epsilon\, U_1(y_*) - \sqrt{D\epsilon}\, b$. We have to check that if $y_*=x_1$ then $y_1\simeq x_*$ up to small error vanishing with $\epsilon$. With $y_*=x_1$, we have
\begin{eqnarray*}
y_1 && = x_1 - \epsilon\, U_1(x_1) - \sqrt{D\epsilon}\, b  \\
&& = (x_*+\epsilon\, U_1(x_*) + \sqrt{D\epsilon}\, b) -\epsilon\, U_1(x_*+\epsilon\, U_1(x_*) + \sqrt{D\epsilon}\, b) - \sqrt{D\epsilon}\,b  \\
&&= x_* +O(\epsilon^{3/2}).
\end{eqnarray*}
We can iterate this procedure all along the trajectory. The number of iterations is of order $1/\epsilon$. Hence, by going up and down using the forward and backward processes, we get back to the initial position up to $O(\epsilon^{1/2})$. Thus, \eqref{eq:discrete-reversed} is indeed a discretisation of the reversed process as well as of \eqref{eq:sde-discret}.

Let us now look at the path integral approach. Let us first write the path integral representation of the forward probability distribution:
\be
\mathbb{P}_\uparrow(x,t|x_0,t_0) = \int_{x_{s=0}=x_0}^{x_{s=t}=x}\hskip -0.7 truecm Dx\quad e^{-\frac{1}{2D}\int_{t_0}^t ds(\dot x_s - U_s(x_s))^2}.
\ee
This integral has to be understood with the Ito convention, in the sense to the cross term in the integral reads
\[ \int_{t_0}^t ds\, \dot x_s\, U(x_s) = \int_{t_0}^t dx_s\, U_s(x_s)\vert_\mathrm{Ito}.\]
This alternatively means that the above path integral is rigorously defined via the Brownian measure as
\be
\mathbb{P}_\uparrow(x,t|x_0,t_0) = \mathbb{E}_{x_0}\big[ e^{\int_{t_0}^t dx_s\, U_s(x_s) -\frac{1}{2D}  \int_{t_0}^t ds\,U_s(x_s)^2}\, \mathbb{I}_{x_t=x}\big],
\ee
with $x_s$ a Brownian normalised to $dx_s^2=D ds$. In the above expression, one recognises the standard exponential martingale, so that the path integral representation of the transition kernel $\mathbb{P}_\uparrow(x,t|x_0,t_0)$ is nothing else than a direct application of the Girsanov's theorem. 

Let us now do a change of variable in the path integral from $x_s$ to $y_s=x_{t-s}$. Naively, the integral $\int_{t_0}^t ds(\dot x_s - U_s(x_s))^2$ is mapped into $\int_{0}^{t-t_0} ds(\dot y_s - \tilde U_s(y_s))^2$ with $\tilde U_s(x) := - U_{t-s}(x)$ as before. However, one has to be a bit more careful because the Ito convention is mapped into the anti-Ito convention under time reversal. Hence, under the transformation $x_s\to y_s=x_{t-s}$, we get
\be
 \int_{t_0}^t ds(\dot x_s - U_s(x_s))^2\vert_\mathrm{Ito} = \int_{0}^{t-t_0} ds(\dot y_s - \tilde U_s(y_s))^2\vert_\mathrm{anti-Ito} .
 \ee
Now, a simple check yields
\be 
\int_{0}^{t-t_0} dy_s \tilde U_s(y_s)\vert_\mathrm{anti-Ito} = \int_{0}^{t-t_0} dy_s \tilde U_s(y_s)\vert_\mathrm{Ito} + \int_{0}^{t-t_0} ds\, \tilde U_s'(y_s).
\ee
Hence,
\be
\frac{1}{2D}\int_{t_0}^t ds(\dot x_s - U_s(x_s))^2\vert_\mathrm{Ito} 
= \frac{1}{2D}\int_{0}^{t-t_0} ds(\dot y_s - \tilde U_s(y_s))^2\vert_{Ito} + \frac{1}{D}  \int_{0}^{t-t_0} ds\, \tilde U_s'(y_s),
\ee
or, reciprocally (by changing the role of $x$ and $y$ as well as $a$ and $\tilde a$),
\be
\frac{1}{2D}\int_{0}^{t-t_0} ds(\dot y_s - \tilde U_s(y_s))^2\vert_{Ito} = \frac{1}{2D}\int_{t_0}^t ds(\dot x_s - U_s(x_s))^2\vert_\mathrm{Ito} + \frac{1}{D}  \int_{t_0}^t ds\, U_s'(x_s).
\ee

We can thus conclude, starting for the path integral representation of the transition kernel of the reversed process, that:
\bea
 \mathbb{P}_\downarrow(x_0,s=t-t_0|x,0) &&=  \int_{y_{0}=x}^{y_{t-t_0}=x_0}\hskip -0.7 truecm Dx\quad e^{-\frac{1}{2D}\int_{0}^{t-t_0} ds(\dot y_s - \tilde U_s(y_s))^2} \\
 &&=  \int_{x_{t_0}=x_0}^{x_{t}=x}\hskip -0.7 truecm Dx\quad e^{-\frac{1}{2D}\int_{t_0}^t ds(\dot x_s - U_s(x_s))^2- \frac{1}{D}  \int_{t_0}^t ds\, U_s'(x_s)}
\eea
A direct application of the Feymann-Kac theorem then proves \eqref{eq:dual-eq}.

\subsection{IIb- Diffusion in a random flow}

Consider the diffusion in the time dependent random flow $\xi_t(x)$. For simplicity of notations
we first consider a single copy. In a second stage, we apply it to each $\xi_t^\pm(x)$, 
and to the reverse process, as in the text.
\be \label{eqx1} 
dx_t = (v + \xi_t(x_t)) dt + \sqrt{D_0} dB_t 
\ee 
The PDF of the position of the particle at time $t$, $P(x,t)= \langle \delta(x_t-x) \rangle_B$ and 
the associated CDF, $\hat P(x,t)=\langle \theta(x_t-x) \rangle_B = \int_x^{+\infty} dy P(y,t)$, satisfy
the Fokker-Planck equation, and its integrated version
\be \label{FPapp}
\partial_t P = \frac{D}{2} \partial_x^2 P - \partial_x [ (v + \xi_t(x)) P(x,t) ] \quad , \quad 
\partial_t \hat P = \frac{D}{2} \partial_x^2 \hat P - (v+ \xi_t(x)) \partial_x \hat P (x,t) 
\ee 
Here $D=D_0$ if $\xi_t(x)$ is smooth in time, but $D=D_0 + \kappa \delta_a(0) \approx D_0 + \kappa/a$ for
the model studied here, where $\xi_t(x)$ is delta-correlated in time, and Ito time-discretization
is used for the random field. This coarse-graining of the diffusion coefficient
is shown as follows. Indeed from \eqref{eqx1} one has, using Ito's rules
\be
d(x_t^2) = 2 x_t dx_t + (dx_t)^2 = 2 x_t dx_t + (\xi_t(x_t) dt)^2 + D_0 dB_t^2 
+ \xi_t(x_t) dt dB_t + O(dt^{3/2}) 
\ee 
and, averaging, $\mathbb{E}[\langle d(x_t^2) \rangle_B ]=  (\kappa \delta_a(0) + D_0) dt + O(dt^{3/2})$.

This problem was studied in \cite{TTPLDDiffusion} and in \cite{BarraquandSticky}.
In \cite{TTPLDDiffusion} the following argument was given, and some of its consequences were
checked numerically. Using the Ito prescription, and averaging \eqref{FPapp} over $\xi_t$, we
see that $\mathbb{E}[P(x,t)]= \frac{1}{\sqrt{2 \pi D t}} e^{- \frac{(x-v t)^2}{2 D t}}$, the free diffusion
with diffusion coefficient $D$. We are interested in looking in the region $x \sim (v+\theta) t$ at large $t$, with $\theta \neq 0$, i.e. away from the typical direction $x=v t$. We are thus probing the tail of the PDF, and we factor out the main dependence in $x$ (which is exponential) in that region by writing
\be
P((v+\theta) t + x,t)= Z(x,t) e^{- \frac{x \theta}{D} - \frac{\theta^2}{2 D} t }
\ee 
Then $Z(x,t)$ satisfies 
\bea \label{she1} 
\partial_t Z(x , t) =  \frac{D}{2} \partial_x^2 Z(x , t)  +  \frac{\theta}{D} \hat \xi(x,t) Z(x,t) - \mu \, \partial_x (\hat \xi(x,t) Z(x,t)) 
\eea  
with $\mu=1$. Here $\hat \xi(x,t)=\xi(x+(v+\theta) t,t)$ is a Gaussian noise which has exactly the
same correlator (hence distribution) as $\xi(x,t)$.
The equation with $\mu=0$ is the stochastic heat equation (SHE), which
is related to the KPZ equation via the Cole-Hopf transform $h(x,t)= \log Z(x,t)$. 
In \cite{TTPLDDiffusion} it was argued that the additional term $\partial_x (\hat \xi Z)$ in \eqref{she1}, since it contains additional derivatives, is irrelevant by power counting above a certain scale. For
$\mu=0$ the KPZ field is known to display scale invariant fluctuations
and we can rescale $h(x,t) = b^{\alpha} \tilde h(x/b, t/b^z)$ with $b$ large and $z=3/2$ and $\alpha=1/2$ the dynamic and roughness exponent of the KPZ class, with $\tilde h=O(1)$. From the scale invariance of the Gaussian white noise, under rescaling the last term in (\ref{she1}) receives an additional
factor $b^{\alpha-1}$ as compared to the first one. This heuristic suggests that the second source of noise is irrelevant as long as $\alpha <1$. The scale above which these terms can be neglected 
is estimated  \cite{TTPLDDiffusion} to be the diffusion scale $x= x_0 = D/\theta$ and $t_0=x_0^2/D=D/\theta^2$. This should be compared to the characteristic length and time scales,
noted $x^{**}$ and $t^{**}$, of the KPZ equation
obtained from \eqref{she1} at $\mu=0$ with $h(x,t)=\log Z(x,t)$
\bea \label{kpz1} 
\partial_t h(x , t) =  \frac{D}{2} \left( \partial_x^2 h(x , t)  + (\partial_x h(x,t))^2 \right)+ \frac{\theta}{D} \hat \xi(x,t) \, .
\eea  
Approximating these scales by the ones obtained for KPZ with white noise ($a\to0$), one obtains \cite{TTPLDDiffusion}, $x^{**}= \frac{D^3}{\kappa \theta^2}$ and $t^{**} =2 D^5/(\kappa^2 \theta^4)$.
The SHE/KPZ equation describe the crossover between the so-called Edwards-Wilkinson (EW) regime
(at scales shorter than $x^{**},t^{**}$) and the KPZ fixed point behavior at large scales. In the EW
regime the non-linear term can be neglected in the KPZ equation, hence the fluctuations of $\log Z$ 
are Gaussian and grow as $t^{1/4}$, with $z=2$. There are thus two cases \cite{TTPLDDiffusion}

(i) moderate $\theta$, in which case $\log P$ shows at large time the fluctuations of a height field in the KPZ class: the additional term $O(\mu)$ in \eqref{she1} is irrelevant, 
but produces sizeable renormalisation of the amplitudes (see below).

(ii) small $\theta \ll \theta^* = D^2/\kappa$ in which case the scale of the KPZ equation 
is $t^{**} > t_0$, and the PDF of $\log P$ is described by then described by the finite time KPZ equation,
above the scales $x_0$ and $t_0$, with small renormalisation of the amplitudes 
(see SM of \cite{TTPLDDiffusion} p. 13-14). In the time window 
$t_0 < t < t^{**}$ these arguments predict that $\log P$ equals $\log \overline{P}$ 
with subdominant Gaussian EW-type fluctuations. A similar property was proved 
for discrete models of walks in \cite{EWforDiffusion}. 

The general picture for diffusion in time-dependent random fields, which emerges from the above qualitative arguments in \cite{TTPLDDiffusion}, as well as from exact results in \cite{BarraquandSticky}
(as discussed in \cite{BarraquandPLD2019}) and unveiled
for the first time in the exact solution of the Beta random walk model on the square lattice \cite{BarraquandCorwinBeta,TTPLDBeta} is that the CDF (or equivalently, the PDF)
in an atypical direction obeys at large time a large deviation principle of the form
\be \label{ld} 
\log {\rm Prob}(x_t > (v + \theta) t) \simeq - J(\frac{\theta}{\theta^*}) \frac{t}{t^*} + 
(\frac{t}{t^*})^{1/3} G(\frac{\theta}{\theta^*}) \chi_2 
\ee 
where $\chi_2$ is a random variable with TW
distribution. The functions $J(\theta)$, $G(\theta)$, as well as the parameters
$t^*$ and $\theta^*$ a priori depend on the details of the model. In Ref. \cite{BarraquandSticky} a continuum
diffusion model formulated in terms of sticky Brownian motions was shown to be
integrable. The generator for the $n$ multipoint moment of the CDF
for this model is of the form $\frac{D}{2} \sum_i \partial_{x_i}^2 + \frac{\kappa}{2} \sum_{i \neq j} 
\delta(x_i-x_j) \partial_{x_i} \partial_{x_j}$, so formally it corresponds to a delta function
correlator of the noise (we write "formally" because the limit $a\to0$ is mathematically very
delicate to define). For that model the parameters in \eqref{ld} are
\be \label{tt} 
t^* = \kappa^2/D^3 \quad , \quad \theta^* = D^2/\kappa
\ee 
and the functions are \cite{BarraquandSticky}
\be J(\theta) = \max_{u>0} [ \frac{1}{2} \psi_2(u) + \theta \psi_1(u) ] \quad , \quad 
G(\theta)= \frac{1}{2^{1/3}} (- \frac{1}{2} \psi_4(s) - \theta \psi_3(s) )^{1/3} \quad , \quad
\theta = - \frac{1}{2} \frac{\psi_3(s)}{\psi_2(s)} 
\ee
and they have the following expansions at small argument
\bea \label{jj} 
J(\theta) = \frac{\theta ^2}{2}+\frac{\theta ^4}{24}-\frac{\theta
   ^6}{80}+\frac{43 \theta ^8}{2688}+O\left(\theta ^{10}\right)
\quad , \quad G(\theta) = \frac{1}{2^{1/3}} \theta^{4/3} \left( 1+\frac{\theta ^4}{16}-\frac{29 \theta^6}{144}+O\left(\theta ^8\right) \right) 
\eea 
For the model studied here (with smooth short range noise correlations), there
is no exact solution but the arguments given in \cite{TTPLDDiffusion} (i.e the approximation using to the KPZ equation) lead in fact to the same parameters \eqref{tt} and to 
$J(\theta) = \frac{\theta ^2}{2}+\frac{\theta ^4}{24}$ and $G(\theta)=\frac{1}{2^{1/3}} \theta^{4/3}$,
i.e they agree with the results of \cite{BarraquandSticky} for small values of $\theta/\theta^*$. 
We call "quadratic approximation" in the text the further approximation of $J(\theta)$ 
by it leading term $J(\theta) \approx \frac{\theta ^2}{2}$. We note that the
higher orders in \eqref{jj} presumably arise as further corrections from the term $O(\mu)$ in \eqref{she1}, 
which cannot be neglected when $\theta/\theta^*$ is not small.

Let us remark that the description in terms of the KPZ equation becomes exact in the so-called
moderate deviation regime $x \sim t^{3/4}$, which corresponds to $\theta \to 0$, as shown recently
\cite{BarraquandPLD2019}.

To connect with the text, the implication of the above analysis is that one can write,
for any fixed $\theta_\pm$ and large $t-t_0$
\be
P^\pm_{t,t_0}(x,(\mp v + \theta_\pm) (t-t_0) + \tilde y) = Z^\pm_{t,t_0}(x,\tilde y)
e^{- \frac{\tilde y}{D} \theta_\pm - \frac{\theta_\pm^2}{2 D} (t-t_0)} 
\ee 
in the region $\tilde y=o(t-t_0)$ (such $|\tilde y| \ll \theta_\pm (t-t_0)$, so that we are
still exploring atypical trajectories). Here the factors
$Z^\pm_{t,t_0}(x,\tilde y)$ satisfies a SHE similar to \eqref{she1} 
(replacing $\tilde y$ by $y$)
\bea \label{she12} 
- \partial_{t_0} Z^\pm_{t,t_0}(x,\tilde y) =  \frac{D}{2} \partial_{\tilde y}^2 Z^\pm_{t,t_0}(x,\tilde y)   \mp  \frac{\theta_\pm}{D} \hat \xi_{t_0}^\pm(\tilde y) Z^\pm_{t,t_0}(x,\tilde y)  \pm  \mu \, 
\partial_x (\hat \xi_{t_0}^\pm(\tilde y) Z^\pm_{t,t_0}(x,\tilde y)) 
\eea  
with $\mu=1$. Here $\hat \xi_{t_0}(\tilde y)=\xi_{t_0}(\tilde y + (\mp v + \theta_\pm) (t-t_0))$ is a Gaussian noise which has exactly the
same correlator (hence distribution) as $\xi_t(\tilde y)$. Note that the noise
in the associated KPZ equations (i.e. Eq. \eqref{she12} with $\mu=0$) depends implicitly on $\theta_\pm$.

Plugging these expressions in \eqref{eq} we obtain a generalization of
Eq. \eqref{eq2} in the text for any value of the directions $\theta_\pm$.
The integration $\int dy^\pm$ in \eqref{eq} is sampled in two steps, first
by sampling over $\theta_\pm$ which amounts to perform increments in $y^\pm$ of sizes $O(t)$,
and then sampling over $\tilde y^\pm$ inside each of these increments. This leads
to the replacement $\int dy^\pm \to t \int d\theta_\pm \int_{|\tilde y^\pm| \lesssim o(t)} d\tilde y^\pm$
in \eqref{eq} and to
\be \label{eq22} 
\! \! \langle e^{-q (S_n-s_n) } \rangle_B \simeq
t^2 \int d\theta_+ d \theta_- \int_{|\tilde y_\pm| \lesssim o(t)} \hskip -0.5 cm d\tilde y^+ d\tilde y^-
\frac{Z^+_{t,t_0}(x,\tilde y^+) Z^-_{t,t_0}(x,\tilde y^-) 
e^{ - \frac{\tilde y^+ \theta_+ + \tilde y^- \theta_-}{D} - \frac{\theta_+^2 +\theta_-^2 }{2D} t }
}{[ \cosh(\frac{\pi (-2 v + \theta_{+} - \theta_{-}) t + 
\tilde y^+-\tilde y^-)}{4 v \tau_0} )] ^{2 q \delta_n}}
\ee 
where from now on we choose implicitly $t_0=0$ to simplify notations.
Note that Eq. \eqref{eq2} in the text amounted to focus on the region $\theta_\pm = \pm v$.
Keeping $\theta_\pm$ as a summation variable samples trajectories at various angles
in the $(x,t)$ plane, 
and integrating over $\tilde y_\pm$ tests the fluctuations of the trajectories
around these angles. 

Let us now estimate the summation over $\theta_\pm$ at large $t$ by a saddle point
approximation over two variables. We will work with the same accuracy as the quadratic 
approximation discussed in the text, valid for $v < \theta^*$. The contribution to the
saddle point equation of the factors 
$Z^+_{t,t_0}(x,\tilde y^+) Z^-_{t,t_0}(x,\tilde y^-)$ is subdominant of order $O(\theta^4 t)$ within
this approximation and focusing only on the terms $O(t)$ we obtain
\bea \label{col2} 
&& \log \langle e^{-q (S_n-s_n) }  \rangle_B  \simeq
- \min_{\theta_\pm} \left[ q \frac{\pi \delta_n}{2 v \tau_0} 
|-2 v + \theta_{+} - \theta_{-})| t 
+ \frac{\theta_+^2 +\theta_-^2 }{2D} t \right] + O(t^{1/3}) 
\\
&& \hskip 1.0 truecm  = - \frac{2 t}{D} \min_{\theta, \sigma} \left[ \frac{q v}{q_c} |\theta+v| + \frac{\theta^2}{2} 
+ \frac{\sigma^2}{2}  \right] + O(t^{1/3}) 
\eea 
where $q_c=\frac{2 v^2 \tau_0}{D \pi \delta_n}$ as in the text, and we have defined the new variables
$\theta= \frac{\theta_- - \theta_+}{2}$ and $\sigma= \frac{\theta_- + \theta_+}{2}$.
The minimum is at $\sigma=0$ hence $\theta_+=- \theta_-$. Hence we recover
exactly the minimisation problem of the text in \eqref{col} in terms of the variable $\theta$,
with the same result (here within the quadratic approximation). This establishes
the link between the formula \eqref{eq} and \eqref{col}. 

Having determined the location of the saddle point in the variables $\theta_\pm$, 
we can return to the evaluation
of the sample to sample fluctuations. For $q>q_c$ the inverse cosh term in \eqref{eq22}
can effectively be replaced
by a delta function at large time, as described in the text, and the fluctuations are given by \eqref{res1}.
For $q<q_c$, the optimal trajectories do not touch. The fluctuating term in
\eqref{eq22} is proportional to 
$\int_{|\tilde y_\pm| \lesssim o(t)}  d\tilde y^+ d\tilde y^-
Z^+_{t,t_0}(x,\tilde y^+) Z^-_{t,t_0}(x,\tilde y^-)$ (note that the term 
exponential $e^{c (y^--y^+)}$ vanishes, $c=0$, from the saddle point condition).
Since the two noises are uncorrelated but now the endpoints $y_\pm$ can fluctuate
freely within an $O(t^{2/3})$ region, we conclude that, for $q<q_c$, $\log \langle e^{-q (S_n-s_n) }  \rangle_B$ fluctuates as
$\propto t^{1/3} \chi$ where $\chi$ is now the sum of two independent GOE TW distributed random variables
(each one associated to the KPZ class with flat initial conditions).

\subsection{IIc- Single trajectory behavior} 

We use the following fact. Consider $X_s$ which satisfies $d X_s = \xi_s(X_{s^-}) ds$ 
(the minus refers to the Ito convention),
where $\xi_s$ is random Gaussian field of zero mean and delta correlated in time,
with $\mathbb{E}(\xi_s(x) \xi_u(y))= \delta(s-u) R(x-y)$.
We argue now
that for a fixed initial condition $X_0=x_0$, $X_s$ is a Brownian. 
To this aim we consider the correlation of its increments. First we show that
they are independent at different times. Indeed, for $u<s$ one has
\bea
&& \mathbb{E}[ \xi_s(X_{s^-}) \xi_u(X_{u^-}) ] = 
\mathbb{E}[  \mathbb{E}[ \xi_s(X_{s^-}) \xi_u(X_{u^-}) | {\cal F}_{s^-} ]] \\
&& = \mathbb{E}[  \xi_u(X_{u^-})  \mathbb{E}[ \xi_s(X_{s^-}) | {\cal F}_{s^-} ]] = 0 
\eea
In the first line we use the Ito prescription (hence time $s^-$ and $u^-$) 
and we condition on the information on the process up to any time $s'<s$ 
(defined by the filtration ${\cal F}_{s^-}$). In the second line we use the fact that
$\xi_u(X_{u^-})$ is non-random variable with respect to the measure conditioned 
on ${\cal F}_{s^-}$ to pull it out from the central expectation. The resulting conditional expectation
vanishes  $\mathbb{E}[ \xi_s(X_{s^-}) | {\cal F}_{s^-} ]=0$ because $\xi_s(\cdot)$ is
independent of the process up to time $s^-$ and has zero mean. 
To treat the equal time correlation one must discretize with infinitesimal
increments, and one finds
\be
\mathbb{E}[ \delta \xi_s(X_{s^-}) \delta \xi_u(X_{u^-}) ] = R(0)\, \delta s 
\ee 
where $\delta \xi_s(x) = \int_s^{s+ \delta s} du\, \xi_u(x)$. 

\subsection{IId- Reflection principle for random trajectories in independent fields $\xi^\pm_t$} 

\begin{figure}
\includegraphics[width=0.25\linewidth]{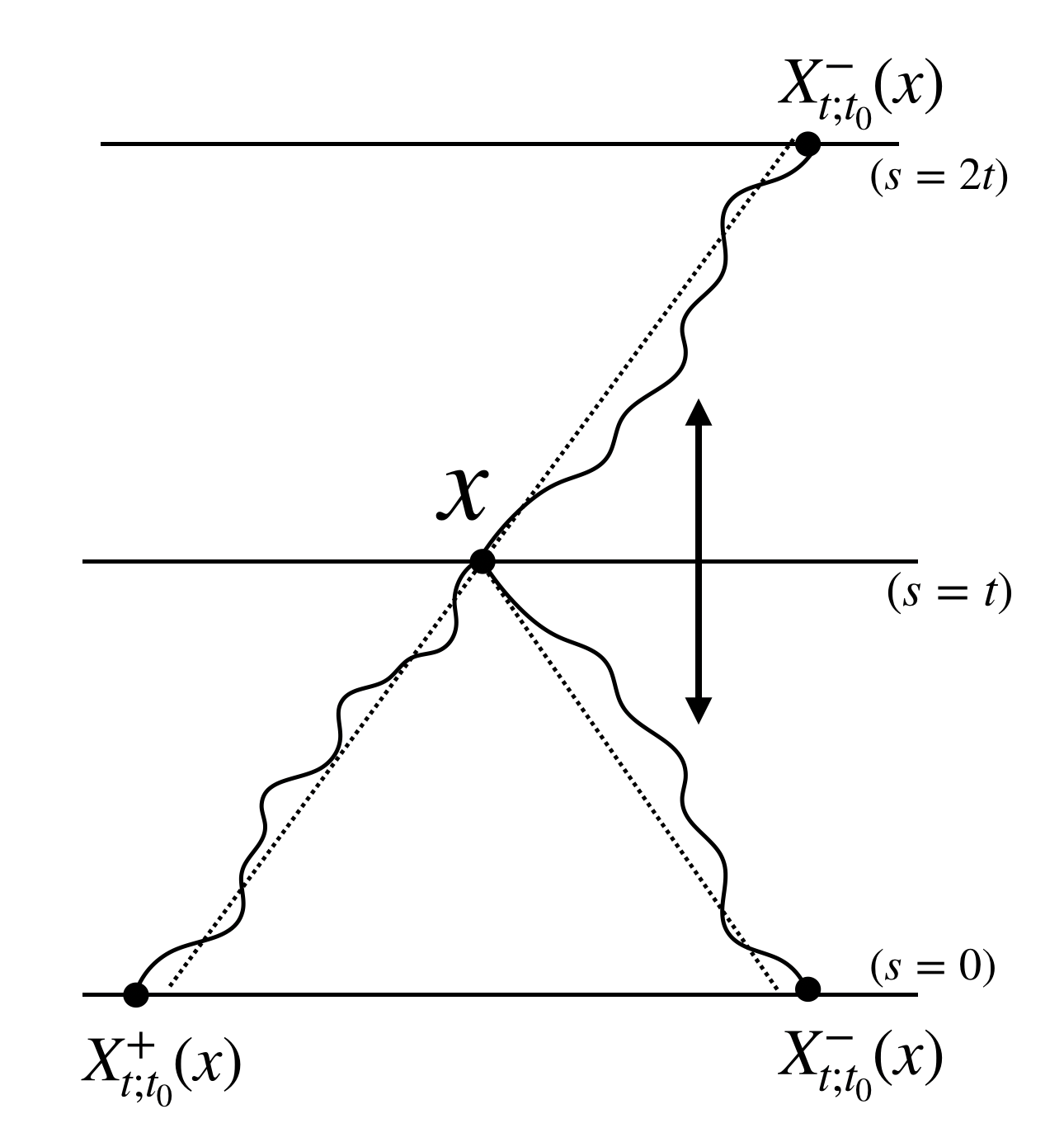}
{\caption{The reflection principle used to construct the process $\tilde x_s$ in Eq. \eqref{langevin2}.
The portion of the curve for $s>t$ is obtained by reflection from the $X^-$ trajectory.}
\label{fig:fig3}}  
\end{figure}

Let us justify the property $|X_t^+(x)-X_t^-(x)| \equiv |\tilde x_{2t} -\tilde x_0|$ mentioned in
the text.
For this
one defines the random fields
$\tilde u_s = v + \tilde \xi_s^+$ for $0<s<t$ and 
$\tilde u_s = v + \tilde \xi_{2t-s}^-$ for $t<s<2 t$.
One defines $\tilde x_s$ as the solution of
\be \label{langevin2} 
\frac{d \tilde x_s}{ds}  =  \tilde u_s(\tilde x_s) + \sqrt{D_0} \frac{d B_s^\pm}{ds} 
\ee
with the condition that $\tilde x_{s=t}=x$ (see Figure \ref{fig:fig3}). By construction
$\tilde x_{s=0}=X_t^+(x)$, and $\tilde x_{s=2t} \equiv X_t^-(x)$.
Hence $\tilde x_{s=2t}  - \tilde x_{s=0} \equiv X_t^-(x) - X_t^+(x)$.

For a fixed $x$, sampling over both the random fields $\xi^\pm$ and the Brownians $B^\pm$,
$\tilde x_s$ as a process is equivalent to $\tilde x_{s} \equiv x+ v (t-s) + \sqrt{D} B(s-t)$,
where $B(s)$ is a unit two sided Brownian motion (with $B(0)=0$). This leads to the "typical" behavior of the
entropy $S_n$ given in the text, with fluctuations $\sim t^{1/2}$
(note that there are corrections to $D$ from the
Jacobian, see section IIe below).

Note that there are however non trivial correlations between the processes
$\tilde x_s$ for different values of $x$ in the same random field configuration.
As mentioned above in Section IIb, in the typical regime,
sample to sample fluctuations are subdominant, i.e. $O(t^{1/4})$, and described by the EW equation 
(for discrete walks this was shown in \cite{EWforDiffusion}). Hence we
similarly expect that, as a function of $x$, the field $S_n =S_n(x,t)$ 
is described by the Edwards-Wilkinson equation with sample to sample
fluctuations scaling as $t^{1/4}$.

\subsection{IIe- Jacobian} \label{app:jacobian}

We now keep track of the Jacobian in \eqref{Sn}, and, after the same steps as in the text
we obtain
\be \label{Gt22} 
e^{(1-n) S_n(t)} \simeq 
 \frac{A_\phi^2 (\frac{\pi}{4 v \tau_0})^{4 \Delta_n} (X^{-\prime}_t(x_0) X^{+\prime}_t(x_0) X^{-\prime}_t(x) X^{+\prime}_t(x))^{\Delta_n}}{[ \cosh( \frac{\pi (X_t^+(x_0)-X_t^-(x_0))}{4 v \tau_0} )
 \cosh( \frac{\pi (X_t^+(x)-X_t^-(x))}{4 v \tau_0})  ]^{2 \Delta_n}}
\ee 
We now show that the Jacobian factors can be evaluated as an integral over the
trajectories of the process $X^\pm_{t,t_0}(x)$, which, as a function of $s=t-t_0$, is the reverse processes
$y^\pm_s=X^\pm_{t,t-s}(x)$ defined in Eq.
\eqref{Xtt} with $\tilde a_s(y)=\mp (v+ \xi^{\pm}_{t-s}(y))$.
Taking a derivative w.r.t $x$ of \eqref{Xtt} we obtain
\be
- \partial_{t_0} X^{\pm \prime}_{t,t_0}(x) = \mp  \xi^{\pm \prime}_{t_0}(X^\pm_{t,t_0}(x))
X^{\pm \prime}_{t,t_0}(x)
\ee 
which, using $X_{t,t}(x)=x$ integrates to,
\bea \label{jac2} 
X^{\pm \prime}_{t,t_0}(x) = e^{\mp \int_{t_0}^t d\tau \, \xi^{\pm \prime}_{\tau}(X^\pm_{t,\tau}(x)) }
= e^{\int_{0}^{t-t_0} ds \, \tilde a^{\pm \prime}_s(y^\pm_s) }
\eea 
The same manipulations
as in the text then lead to
\be \label{eqA} 
\langle e^{-q S_n } \rangle_B \simeq \tau_0^{-4 q \delta_n} 
\int dy^+ dy^- \frac{Q^+_{t,t_0}(x,y^+) Q^-_{t,t_0}(x,y^-) }{\cosh(\frac{\pi (y^+-y^-)}{4 v \tau_0} )^{2 q \delta_n}} 
\ee
where $\delta_n = \frac{\Delta_n}{n-1}= \frac{c}{12} \frac{n+1}{n}$. We have defined
\be
Q^\pm_{t,t_0}(x,y) = \langle \delta(y_{t-t_0}^\pm - y) e^{q \delta_n \int_{0}^{t-t_0} ds \, \tilde a^{\pm\prime}_s(y^\pm_s)} 
\rangle_{\tilde B^\pm} 
\ee 
where the average is
over the Brownian and the reverse processes $y^\pm_s=X^\pm_{t,t-s}(x)$, 
with $y^\pm_0=x$. The $Q^\pm$'s now replace the $P^\pm$ in the formula of the text.
We can use now a path integral representation (see previous Section), and obtain
\bea
Q^\pm_{t,t_0}(x,y) =   \int_{y^\pm_{s=0}=x}^{y^\pm_{s=t-t_0}=y} \, Dy_s \, \, e^{- \int_{0}^{t-t_0} ds
[ \frac{1}{2D} (\dot y_s - \tilde a^\pm_s(y^\pm_s))^2 - q \delta_n  \tilde a^{\pm\prime}_s(y^\pm_s)]} 
\eea
One has
\be \label{FPn}
- \partial_{t_0} Q^\pm_{t,t_0}(x,y)  = [\frac{D}{2} \partial_y^2 \pm \partial_y (v+\xi^\pm_{t_0}(y) )
 \mp q \delta_n  \xi^{\pm \prime}_{t_0}(y) ]
Q^\pm_{t,t_0}(x,y)
\ee 
Until now \eqref{eqA} with $Q^\pm_{t,t_0}(x,y)$ satisfying \eqref{FPn} are exact. The latter can be viewed
as an application of Feynman-Kac.

Now, one argues that the last term in \eqref{FPn}, upon the change of variable 
$Z^\pm=Z^\pm_{t,t_0}(x,y)=Q^\pm_{t,t_0}(x,y) e^{\pm \frac{(y-x) v}{D} + \frac{v^2}{2 D} (t-t_0)}$
yields additional contributions in the equation for $Z^\pm$ (see \eqref{she1})
which can be argued to be irrelevant by similar arguments as given around
Eq. \eqref{she1}. The conclusion is that the Jacobian cannot affect the KPZ class behavior
at large scale (up to possible renormalization of the prefactors $c_i$).\\

We can estimate the effect of the Jacobian in the regime of typical fluctuations for the entropy.
Formula \eqref{typ}, taking the Jacobian into account, reads
\bea \label{typ2} 
&& S_n = s_n + 2 \delta_n \log \cosh(\frac{\pi (X_t^+(x)-X_t^-(x))}{4 v \tau_0}) 
+ \delta_n \log[ X^{-\prime}_t(x) X^{+\prime}_t(x) ]
\eea 
where again we have discarded the terms containing $x_0$ in the limit $x_0 \to - \infty$
and we recall that $X_t^\pm(x)=X_{t,0}^\pm(x)$.
The last term can be rewritten, using \eqref{jac2}
\be \label{jac3}
\delta_n \log[ X^{-\prime}_t(x) X^{+\prime}_t(x) ] =
\delta_n  \int_{0}^t d\tau \, \xi^{- \prime}_{\tau}(X^-_{t,\tau}(x))
- \delta_n \int_{0}^t d\tau \, \xi^{+ \prime}_{\tau}(X^+_{t,\tau}(x)) )
\ee 
In the typical regime, $X^\pm_{t,\tau}(x) \simeq \mp (t-\tau) v + o(t)$
and we see that the two terms in \eqref{jac3} behave as two independent Gaussian noises.

Equivalently we can make an exact calculation sampling both $\xi^\pm$ and $B^\pm$.
We also approximate the $\log \cosh$ in \eqref{typ2} by a linear function, which is valid
for $t \gg \tau_0$, so that
\be
S_n \simeq s_n + \frac{\pi \delta_n}{2 v \tau_0} (X_t^+(x)-X_t^-(x)) + 
\delta_n  \int_{0}^t d\tau \, \xi^{- \prime}_{\tau}(X^-_{t,\tau}(x))
- \delta_n \int_{0}^t d\tau \, \xi^{+ \prime}_{\tau}(X^+_{t,\tau}(x)) )
\ee 
It is then easy to see that w.r.t. sampling both $\xi^\pm$ and $B^\pm$, the process
$S_n(t)$ is a Brownian motion in time $t$. Indeed $\xi_t(X)$ and $\xi_t'(X)$ 
are independent. Using similar arguments as in Section we obtain its diffusion coefficient as
\be
\mathbb{E}( \langle S_n(t) S_n(t') \rangle ) - 
\mathbb{E}( \langle S_n(t) \rangle ) \mathbb{E}(\langle S_n(t') \rangle ) 
= D_s \min(t,t')  \quad , \quad D_s = 2 \delta_n^2 
\left( (\frac{\pi}{2 v \tau_0})^2 ( D_0  + \kappa \delta_a(0) ) +  \kappa |\delta''_a(0)| \right)
\ee 
with $\mathbb{E}( \langle S_n(t) \rangle ) =  \pi \delta_n t/\tau_0$ is the typical 
value of the entanglement entropy at time $t$. 

\subsection{IIf- TT correlations}

From the solution \eqref{TT} of the equation of motion for the energy-momentum tensor 
we obtain, by similar methods as above, the two space time points connected correlation
of $T(x,t)$ as 
\bea \label{tt} 
&& C(x_1,t_1;x_2,t_2)= \langle \Psi_0|T(x_1,t_1) T(x_2,t_2) |\Psi_0 \rangle^c 
= (X^{+\prime}_{t_1}(x_1))^2 (X^{+\prime}_{t_2}(x_2))^2 \frac{c/2}{(\cosh( \frac{\pi}{4 v \tau_0} (X^+_{t_2}(x_2) 
- X^+_{t_1}(x_1)))^4} \nonumber 
\eea 
Note that the term containing the Schwarzian derivative drops out since it is a connected
correlation. Since it is a chiral operator, only $X^+_t(x)$ enters the formula.
Since both $X^+_{t_1}(x_1)$ and $X^+_{t_2}(x_2)$ enter the formula, and both feel the
random field $\xi^+$, we need
to study now two trajectories in the same environment. This case is more
difficult, and we leave the detailed analysis to a future work. Let us
only sketch some features of the simplest case of equal time correlation, $t_1=t_2=t$.
In the region $x_2-x_1 \sim v t$ one can neglect the correlations between
the environments seen by the two diffusions, and one can estimate the moments
$\langle C^q \rangle_{B^+}$ at large $t$, where $C=C(x_1,t;x_2,t)$, by a variational method similar to the one in Section IIb. Within the quadratic approximation, we find two cases: (i) if
$\frac{x_2-x_1}{2 t} > \frac{D \pi q}{v \tau_0}$ the optimal trajectories 
do not touch and $\log \langle C^q \rangle_{B^+} \sim - \frac{\pi q}{v \tau_0} (2 \frac{x_2-x_1}{2 t}  - 
\frac{D \pi q}{v \tau_0}) t + o(t)$, (ii) if $\frac{x_2-x_1}{2 t} < \frac{D \pi q}{v \tau_0}$, the
inverse cosh function in \eqref{tt} can be replaced by a delta function, and the optimal trajectories
meet near their endpoint, with $\log \langle C^q \rangle_{B^+} \sim - \frac{(x_2-x_1)^2}{4 D t} + o(t)$.
In both cases $\log \langle C^q \rangle_{B^+}$ exhibits $O(t^{1/3})$ sample to sample fluctuations related to Tracy-Widom distributions, as described in previous sections. In the region $x_2-x_1 = O(t^{2/3})$ one must take into account 
that the two optimal trajectories see the same environment, and stick over a finite fraction of
their length. The fluctuations are then of the same type as 
$\log[\int dy Z^+_{t,t_0}(x_1,y) Z^+_{t,t_0}(x_2,y)]$, i.e. as the two point function
of the KPZ equation with flat initial condition in the large time limit, described by the so-called Airy$_1$
process as a function of $x_2-x_1$.

\end{widetext} 



\begin{thebibliography}{}
\bibitem{footnote0}
But with more singular environments than the ones we consider in the present paper.
\bibitem{footnote6} 
A more local version of the model amounts to replace the Brownian noise in \eqref{eq2} 
by space dependent noise, $B_t^\pm \to \zeta_t^\pm(x)$, independent of
$\xi_t^\pm(x)$. Interestingly, the results are unchanged as long as one restricts to
observables involving at most one trajectory $X_t^\pm(x)$ at one point,
such as $S_n$ (in the limit $x_0 \to -\infty$), or \eqref{single}. They deviate
however for correlations such as \eqref{TT0}.
\bibitem{footnote01}
With appropriate sum over conformal blocks (if needed).
\bibitem{footnote3} 
We substract to the entropy the contribution coming from the endpoint $x_0$.
Taking the limit $x_0 \to - \infty$ first, amounts to consider time $t$ which can 
be large but still much shorter than the interval length $x-x_0$.
\bibitem{footnote4}
Note that the probability for the equality for $S_n$ is expected to follow
the same large deviation law.
\bibitem{footnote5} In \eqref{col} to obtain $\theta_c$ we used $|\theta+v|=\theta+v$. There is another
saddle point for the choice $|\theta+v|=-(\theta+v)$, $\theta_c=\frac{q}{q_c} v$
for $q<-q_c$. However its contribution is subdominant.
\end{thebibliography}
\end{document}